\documentclass[twoside]{article}
\usepackage{myqic,epsfig}

\usepackage{amssymb}
\usepackage{amsmath}
\usepackage{graphicx}
\usepackage{psfrag}
\usepackage{color}
\usepackage[all,knot,poly]{xy}
\usepackage{youngtab}

\DeclareMathOperator{\Tr}{Tr}
\DeclareMathOperator{\tr}{tr}

\DeclareMathOperator{\E}{\mathbb{E}}

\begin{document}
\setlength{\textheight}{8.0truein}    

\runninghead{The Jones Polynomial: Quantum Algorithms and Applications in Quantum Complexity Theory}
            {Pawel Wocjan and Jon Yard}

\normalsize\textlineskip
\thispagestyle{empty}
\setcounter{page}{1}

\copyrightheading{0}{0}{2003}{000--000}

\vspace*{0.88truein}

\alphfootnote

\fpage{1}

\centerline{\bf
THE JONES POLYNOMIAL: QUANTUM ALGORITHMS AND APPLICATIONS}
\vspace*{0.035truein}
\centerline{\bf IN QUANTUM COMPLEXITY THEORY}
\vspace*{0.37truein}
\centerline{\footnotesize
PAWEL WOCJAN\footnote{School of Electrical Engineering and Computer Science, University of Central Florida, Orlando, FL 32816, USA.  Electronic address: {\tt wocjan@cs.ucf.edu}}}
\vspace*{0.015truein}
\centerline{\footnotesize\it Institute for Quantum Information, Caltech}
\baselineskip=10pt
\centerline{\footnotesize\it Pasadena, CA 91125, USA}
\vspace*{10pt}
\centerline{\footnotesize 
JON YARD\footnote{Electronic address: {\tt jtyard@caltech.edu}}}
\vspace*{0.015truein}
\centerline{\footnotesize\it Institute for Quantum Information, Caltech}
\baselineskip=10pt
\centerline{\footnotesize\it Pasadena, CA 91125, USA}
\vspace*{0.225truein}
\publisher{(June 20, 2006)}{(revised date)}

\vspace*{0.21truein}

\abstracts{
We analyze relationships between quantum computation and a family of generalizations of the Jones polynomial.  Extending recent work by Aharonov et al., we give efficient quantum circuits for implementing the unitary Jones-Wenzl representations of the braid group.  We use these to provide new quantum algorithms for approximately evaluating a family of specializations of the HOMFLYPT two-variable polynomial of trace closures of braids.  We also give algorithms for approximating the Jones polynomial of a general class of closures of braids at roots of unity.  Next we provide a self-contained proof of a result of Freedman et al.\ that any quantum computation can be replaced by an additive approximation of the Jones polynomial, evaluated at almost any primitive root of unity.    Our proof encodes two-qubit unitaries into the rectangular representation of the eight-strand braid group.  We then give QCMA-complete and PSPACE-complete problems which are based on braids.  We conclude with direct proofs that evaluating the Jones polynomial of the plat closure at most primitive roots of unity is a \#P-hard problem, while learning its most significant bit is PP-hard, circumventing the usual route through the Tutte polynomial and graph coloring.
}{}{}

\vspace*{10pt}

\vspace*{3pt}
\communicate{to be filled by the Editorial}

\vspace*{1pt}\textlineskip    

\def\norm#1{ {|\hspace{-.022in}|#1|\hspace{-.022in}|} }
\def\Norm#1{ {\big|\hspace{-.022in}\big| #1 \big|\hspace{-.022in}\big|} }
\def\NOrm#1{ {\Big|\hspace{-.022in}\Big| #1 \Big|\hspace{-.022in}\Big|} }
\def\NORM#1{ {\left|\hspace{-.022in}\left| #1 \right|\hspace{-.022in}\right|} }
\def\nn{\nonumber}
\def\qedsymbol{\rule{7pt}{7pt}}
\def\supp{ {\rm{supp \,}}}
\def\dist{ {\rm{dist }}}
\def\dim{ {\rm{dim \,}}}
\def\oti{{\otimes}}
\def\bra#1{{\langle #1 |  }}
\def\lb{ \left[ }
\def\rb{ \right]  }
\def\tilde{\widetilde}
\def\bar{\overline}
\def\*{\star}
\def\({\left(}		\def\BL{\Bigr(}
\def\){\right)}		\def\BR{\Bigr)}
	\def\BBL{\lb}
	\def\BBR{\rb}
%

\def\1{{\mathbf{1} }}

\def\bb{{\bar{b} }}
\def\ab{{\bar{a} }}
\def\zb{{\bar{z} }}
\def\zbar{{\bar{z} }}
\def\inv#1{{1 \over #1}}
\def\half{{1 \over 2}}
\def\d{\partial}
\def\der#1{{\partial \over \partial #1}}
\def\dd#1#2{{\partial #1 \over \partial #2}}
\def\vev#1{\langle #1 \rangle}
\def\ket#1{ | #1 \rangle}
\def\rvac{\hbox{$\vert 0\rangle$}}
\def\lvac{\hbox{$\langle 0 \vert $}}
\def\2pi{\hbox{$2\pi i$}}
\def\e#1{{\rm e}^{^{\textstyle #1}}}
\def\grad#1{\,\nabla\!_{{#1}}\,}
\def\dsl{\raise.15ex\hbox{/}\kern-.57em\partial}
\def\Dsl{\,\raise.15ex\hbox{/}\mkern-.13.5mu D}
\def\b#1{\mathbf{#1}}
\newcommand{\proj}[1]{\ket{#1}\bra{#1}}
\def\braket#1#2{\langle #1 | #2 \rangle}
%
%
\def\th{\theta}		\def\Th{\Theta}
\def\ga{\gamma}		\def\Ga{\Gamma}
\def\be{\beta}
\def\al{\alpha}
\def\ep{\epsilon}
\def\vep{\varepsilon}
\def\la{\lambda}	\def\La{\Lambda}
\def\de{\delta}		\def\De{\Delta}
\def\om{\omega}		\def\Om{\Omega}
\def\sig{\sigma}	\def\Sig{\Sigma}
\def\vphi{\varphi}
%
%
\def\CA{{\cal A}}	\def\CB{{\cal B}}	\def\CC{{\cal C}}
\def\CD{{\cal D}}	\def\CE{{\cal E}}	\def\CF{{\cal F}}
\def\CG{{\cal G}}	\def\CH{{\cal H}}	\def\CI{{\cal I}}
\def\CJ{{\cal J}}	\def\CK{{\cal K}}	\def\CL{{\cal L}}

\def\CM{{\cal M}}	\def\CN{{\cal N}}	\def\CO{{\cal O}}
\def\CP{{\cal P}}	\def\CQ{{\cal Q}}	\def\CR{{\cal R}}
\def\CS{{\cal S}}	\def\CT{{\cal T}}	\def\CU{{\cal U}}
\def\CV{{\cal V}}	\def\CW{{\cal W}}	\def\CX{{\cal X}}
\def\CY{{\cal Y}}	\def\CZ{{\cal Z}}


\def\bA{\mathbb{A}}	\def\bB{\mathbb{B}}	\def\bC{\mathbb{C}}
\def\bD{\mathbb{D}}	\def\bE{\mathbb{E}}	\def\bF{\mathbb{F}}
\def\bG{\mathbb{G}}	\def\bH{\mathbb{H}}	\def\bI{\mathbb{I}}
\def\bJ{\mathbb{J}}	\def\bK{\mathbb{K}}	\def\bL{\mathbb{L}}
\def\bM{\mathbb{M}}	\def\bN{\mathbb{N}}	\def\bO{\mathbb{O}}
\def\bP{\mathbb{P}}	\def\bQ{\mathbb{Q}}	\def\bR{\mathbb{R}}
\def\bS{\mathbb{S}}	\def\bT{\mathbb{T}}	\def\bU{\mathbb{U}}
\def\bV{\mathbb{V}}	\def\bW{\mathbb{W}}	\def\bX{\mathbb{X}}
\def\bY{\mathbb{Y}}	\def\bZ{\mathbb{Z}}

\def\rvac{\hbox{$\vert 0\rangle$}}
\def\lvac{\hbox{$\langle 0 \vert $}}
\def\comm#1#2{ \BBL\ #1\ ,\ #2 \BBR }
\def\2pi{\hbox{$2\pi i$}}
\def\e#1{{\rm e}^{^{\textstyle #1}}}
\def\grad#1{\,\nabla\!_{{#1}}\,}
\def\dsl{\raise.15ex\hbox{/}\kern-.57em\partial}
\def\Dsl{\,\raise.15ex\hbox{/}\mkern-.13.5mu D}
\def\beq{\begin {equation}}
\def\eeq{\end {equation}}
\def\to{\rightarrow}
\def\h#1{\widehat{#1}}

\def\diag{\mbox{diag}}

\def\isom{\simeq}
\def\alg#1{\left\langle\hspace{-.09cm}\left\langle #1 \right\rangle\hspace{-.09cm}\right\rangle }
\def\argmax{\mbox{argmax}}
\definecolor{gray}{gray}{.8}
\def\com#1{\vspace{.1in}\fcolorbox{black}{gray}{\begin{minipage}{5.5in}#1\end{minipage}}\vspace{.1in}}

\def\Bra#1{{\big\langle #1 \big|  }}
\def\Ket#1{{\big| #1 \big\rangle }}
\def\BRA#1{{\left\langle #1 \right|  }}
\def\KET#1{{\left| #1 \right\rangle }}
\newcommand{\Proj}[1]{\Ket{#1}\Bra{#1}}
\def\Braket#1#2{\big\langle #1 \big| #2 \big\rangle}
\def\pmat#1{\begin{pmatrix} #1 \end{pmatrix}}
\def\poly{\text{\rm poly}}
\def\polylog{\text{\rm poly\! log}}


\def\r{\vtwist, c+(2,1)}
\def\l{\vtwistneg, c+(2,1)}
\def\s{\POS c+(0,0) *{};p-(0,1) *{} **\dir{-};c+(1,0)}
\def\N{\POS "A"-(0,1)="A";"A"}
\def\braid#1#2{\,\,\xy 0;<.4cm,0cm>:<0cm,.5cm>:: \POS (0,#1)="A";"A" #2 \endxy\,\,}
\def\caps#1#2{\POS c="B";"B"+(#1,0),\vcap[#2];"B";"B"}
\def\cups#1#2{\POS c="B";"B"+(#1,0),\vcap[-#2];"B";"B"}

\newtheorem{theo}{Theorem}[section]
\newtheorem{lem}[theo]{Lemma}
\newtheorem{prop}[theo]{Proposition}
\newtheorem{cor}[theo]{Corollary}
\newtheorem{conj}[theo]{Conjecture}
\newtheorem{keyobs}[theo]{Key Observation}

\newtheorem{dfn}[theo]{Definition}
\newtheorem{ex}[theo]{Example}
\newtheorem{problem}{Problem}[section]

\newcommand{\Section}[1]{\section{#1} \setcounter{figure}{0} \setcounter{equation}{0}}
\renewcommand{\thefigure}{\thesection.\arabic{figure}}
\renewcommand{\theequation}{\thesection.\arabic{equation}}
\renewenvironment{proof}[1][Proof]{{\bf #1: }}{$\Box$\\}

\Section{Introduction} \label{section:introduction}
There is evidence that a computer which could manipulate quantum
mechanical degrees of freedom would be more powerful than a classical
computer.  Since the discovery by Shor that a quantum computer could
efficiently factor composite integers, a task which is believed to be
hard on conventional computers, much effort has been expended toward
understanding the capabilities and limitations of quantum computers.  
In this paper, we investigate ways in which the Jones polynomial invariant of knots and links, together with its generalizations, contribute to this understanding.  Witten discovered \cite{witten} that the Jones polynomial could be understood via tools from topological quantum field theories.  Freedman, Kitaev, Larsen and Wang \cite{fklw} established its connection to computer science in the context of topological quantum computation.  In a topological quantum computer, 
the trajectories of particles which are restricted to a plane are braided in order to manipulate the internal state of the computer.  They showed that such computers can be simulated on conventional quantum computers, and also that when the underlying physics of a computer is described by a suitable topological quantum field theory, that it is possible to achieve universal quantum computation. 
 
In \cite{fkw}, it was shown that topological quantum computers could be efficiently simulated by computers based on the standard quantum circuit model, implying the existence of an efficient
algorithm for approximating the Jones polynomial of certain links
obtained from braids.  More recently, Aharonov, Jones and Landau \cite{ajl} gave explicit quantum algorithms for approximating the Jones polynomial of either the trace or plat closure of a braid on a quantum computer.  Our first contribution in this paper is a family of quantum circuits implementing the so-called Jones-Wenzl unitary representations of the braid group, generalizing the circuits for the Jones representation from \cite{ajl}.
We then show how to use our circuits to approximate a family of evaluations of the HOMFLYPT two-variable polynomial.  We also sketch two different quantum algorithms for approximating the Jones polynomial of links obtained via a class of closures generalizing the trace and plat closures.   In the sense of \cite{bflw}, our quantum algorithms obtain additive approximations of Jones and HOMFLYPT polynomial evaluations in polynomial-time.

In \cite{flw1,flw2}, a converse to the above simulation results was proved.    
More specifically, it was shown there that quantum circuits can be simulated by braids, in such a way that the output probabilities of the quantum circuit are functions of the Jones polynomial of the plat closure of some braid of comparable length to that of the original circuit.  Our third contribution is a simpler proof of their main result.  In fact, this work grew from our attempts to understand the connections between these works and  \cite{ajl}.

Finally, we give applications of the Jones polynomial to quantum complexity theory.  We begin by restating a result that was proved in \cite{bflw}, showing that a machine which obtains an additive approximation of the Jones polynomial of the plat closure of a braid is equally as powerful as a quantum computer.  
We then introduce two new problems, \textbf{Increase Jones Plat} and \textbf{Approximate Concatenated Jones Plat}.  The first asks if a given braid can be conjugated by another braid from a given class such that the Jones polynomial of its plat closure is nearly
maximal.  We prove that this problem is complete for the
complexity class QCMA, a certain quantum analog of NP.  The latter problem asks if a given braid, after being concatenated with itself exponentially many times, has a large plat closure.  We show that this problem is PSPACE-complete.
Finally, we give self-contained proofs that learning $n$ of the most significant bits of the Jones polynomial of the plat closure of an $n$-strand braid is a \#P-hard problem, while learning its most significant bit is PP-hard.  This constitutes a simpler, quantum-based proof of the known result that computing the Jones polynomial is \#P-hard.  The original proof of this (see e.g.\ \cite{welsh}) relates the Jones polynomial to the  Tutte polynomial of a signed graph to the chromatic number of that graph to the \#P-complete problem \#SAT.  Our proof, however, shows that learning a linear number of the highest order bits of the Jones polynomial of the plat closure of a braid is \#P-hard.

The paper is organized as follows.  In the next section, we review elements of the theory of links, braids and the Jones polynomial.  There, we review some different ways of turning a braid into a link, while introducing a new type of closure which we call a generalized closure.  We also give precise statements of the approximations achieved by the algorithms.  In Section~\ref{section:reps}, we describe the unitary representations of the braid group used in this paper, highlighting their origins from Hecke algebras.  In Section~\ref{section:localqubitmodel}, we show how to efficiently implement these representations on the state space of a quantum computer.  Section~\ref{section:repformulae} contains derivations of representation-theoretic formulae for the HOMFLYPT and Jones polynomials.
Section~\ref{section:algorithms} shows how to use the quantum circuits of the previous section to construct quantum algorithms for approximating certain evaluations of the HOMFLYPT and Jones polynomials of closed braids.  In Section~\ref{section:encoding}, we provide a self-contained proof of the converse of the results in the previous section -- that local quantum circuits can be simulated by braids.  Finally, in Section~\ref{section:complexity}, we review notions of classical and quantum complexity theory, after which we present our complexity-theoretic results.  We conclude with a discussion in Section~\ref{section:conclusions}.

%

\Section{Background on link invariants and additive approximations}\label{section:links}
Here we give background material on the links invariants used in this paper.  We also formalize the notion of approximation for our quantum approximation algorithms.  We refer the reader to, for example, \cite{welsh} for further details on link invariants.  
A \emph{knot} is a closed, nonintersecting curve in
$\bR^3$.  More generally, a \emph{link} is an embedding of some finite
number of nonintersecting closed curves in $\bR^3$.  Links are identified up to
$\emph{isotopy}$, which means that two links which are related by some
bijection $f\colon \bR^3\to \bR^3$ of the ambient space with itself
for which $f$ and $f^{-1}$ are continuous are considered to be
\emph{equivalent}.  An \emph{oriented link} is a link in which every
component is assigned an orientation.  If $\overrightarrow{L}$ is an
oriented link, we will denote the unoriented link resulting from
forgetting the orientation as $L$.

A central problem in knot theory is to determine, given descriptions
of two links, whether or not they are equivalent.  In order to solve
such a problem, one must decide on how the links are to be described.
One such way is in terms of a \emph{link projection}.  Informally, this is a two-dimensional diagram which uniquely specifies the link up to isotopy.  
The reader is referred to textbooks such as \cite{lickorish, murasugi, ohtsuki,welsh} for examples and further exposition. 
The key property of any link projection is that at each crossing, it keeps track of which string goes above or below the other.
Usually, this information is conveyed
by leaving a gap in the string of the undercrossing.
A classic result in knot theory states that two unoriented links are equivalent
if and only if the link projection of one can be transformed into that of
the other by a finite sequence of \emph{Reidemeister moves} (see e.g.\ any of the texts listed above). 

%
%
\subsection*{The HOMFLYPT and Jones polynomials}
Another way of (partially) distinguishing links involves assigning an
\emph{invariant} to each link so that links which are equivalent up to
isotopy have the same invariant.   
\begin{figure}
\[
\xy
(6,6)*{}="tl";
(-6,6)*{}="tr";
(6,-6)*{}="bl";
(-6,-6)*{}="br";
{\ar|{\hole \; } "tr";"bl"};
{\ar "tl";"br"};
(0,-10)*{L_{+}};
\endxy
\qquad \qquad
\xy
(-6,6)*{}="tl";
(6,6)*{}="tr";
(-6,-6)*{}="bl";
(6,-6)*{}="br";
{\ar|{\hole \; } "tr";"bl"};
{\ar "tl";"br"};
(0,-10)*{L_{-}};
\endxy
\qquad\qquad 
 \xy
(-6,6)*{}="tl";
(6,6)*{}="tr";
(-6,-6)*{}="bl";
(6,-6)*{}="br";
"tl";"bl" **\crv{(-1,0)}?(1)*\dir{>};
"tr";"br" **\crv{(1,0)}?(1)*\dir{>};
(0,-10)*{L_{0}};
\endxy
\]
\fcaption{\label{figure:stuffJonesSkein}$L_+$, $L_-$, $L_0$ denote three oriented links that differ
in a small region, as symbolized by the above diagrams, but are
otherwise the same. We say that the crossing corresponding to $L_+$ is
a positive crossing, and the one corresponding to $L_-$ is a negative
crossing.}
\end{figure}
The \emph{HOMFLYPT polynomial} \cite{homfly,pt} $H_{\overrightarrow{L}}(t,x) \in \bZ[t^{\pm 1},x^{\pm 1}]$ is one such invariant.  It is inductively defined by the \emph{skein relation}
\begin{equation}
t H_{L_{+}}(t,x) - t^{-1} H_{L_{-}}(t,x) = 
x H_{L_{0}}(t,x) \label{eqn:HOMFLYPTskein}
\end{equation}
(see Figure~\ref{figure:stuffJonesSkein}) together with the fact that the HOMFLYPT of the trivial knot (that is, a single unknotted circle) is unity.  
In this paper, we are concerned with various one-variable specializations of this invariant.  
For each integer $k\geq 2$ there is a corresponding \emph{one-variable HOMFLYPT} polynomial in $\bZ[q^{\pm 1/2}]$:
\begin{equation}
H^{(k)}_{\overrightarrow{L}}(q) = H_{\overrightarrow{L}}(q^{k/2},q^{1/2} - q^{-1/2})
\label{eqn:1varHOMFLYPTdefn}
\end{equation}  
This is also known as the $\mathfrak{sl}_k$ invariant.   Restricting to $k=2$ yields the famous  \emph{Jones polynomial}:
\[J_{\overrightarrow{L}}(q) \equiv H^{(2)}_{\overrightarrow{L}}(q) = H_{\overrightarrow{L}}(q,q^{1/2} - q^{-1/2}).\]
If $\overrightarrow{L}$ consists of $m$ unlinked copies of the trivial knot, setting $t = q^{k/2}$ and $x = q^{1/2} - q^{-1/2}$ in (\ref{eqn:HOMFLYPTskein}) implies that  
\[H^{(k)}_{\overrightarrow{L}}(q) = \left(\frac{q^{k/2} - q^{-k/2}}{q^{1/2} - q^{-1/2}}\right)^{m-1}.\]
The exponentiated quantity on the right is known as a \emph{quantum integer} and is easily shown to satisfy 
\begin{equation}
\frac{q^{k/2} - q^{-k/2}}{q^{1/2} - q^{-1/2}} = q^{\frac{k-1}2} + q^{\frac{k-1}{2} - 1} + \cdots + q^{-\frac{k-1}{2}} \label{eqn:qintidentity}
\end{equation}
if $q \neq 1$.  By continuity, we consider this expression to hold as $q\to 1$, recovering the ordinary integer $k$.  
In this paper, we will only be interested in evaluations of these invariants at primitive roots of unity $q = e^{2\pi i /\ell}$, where $\ell$ is a nonnegative integer.  To this end, we define for each integer $k$ and each nonnegative integer $\ell$, the constant\footnote{Our definition here departs from the usual convention \cite{wenzl,flw1} which defines $[k]_q =
\frac{q^{k/2} - q^{-k/2}}{q^{1/2} - q^{-1/2}}$ for arbitrary $q$.  This expression evaluates to ours when $q$ is a primitive $\ell$'th root of unity, which is the case of interest to us.  We warn that other authors (e.g.\ \cite{jones87}) use a different convention for the quantum integer and $q$.}
\[{[k]}_\ell = \left.\frac{q^{k/2} - q^{-k/2}}{q^{1/2} - q^{-1/2}}
\right|_{q=e^{2\pi i/\ell}} = \frac{\sin(\pi k/\ell)}{\sin(\pi/\ell)}.\]
A value which will occur frequently, especially when analyzing the Jones polynomial, is 
\begin{equation}
{[2]}_\ell = \left.q^{1/2} + q^{-1/2}\right|_{q=e^{2\pi i/\ell}} = 2\cos(\pi/\ell). \label{D}
\end{equation}
The Jones polynomial will be the central object of our study, although we will also provide algorithms for approximating the one-variable HOMFLYPT polynomials.    Where convenient, we sometimes write $J\big(\overrightarrow{L},q\big) \equiv J_{\overrightarrow{L}}(q)$.

%
%
\subsection*{The braid group}
The braid group $B_n$ on $n$ strands is generated by counterclockwise twists $\{\sig_1,\sig_2,\dotsc,\sig_{n-1}\}$ which satisfy the relations
\begin{eqnarray}
\sig_i\sig_{j}\sig_i &=& \sig_{j}\sig_i\sig_{j}, \hspace{.2in} |i-j| = 1 \label{braid1}\\
\sig_i \sig_j &=& \sig_j\sig_i \hspace{.4in} |i-j| > 1. \label{braid2}
\end{eqnarray}
The reader should picture $n$ hanging pieces of string, and interpret
each generator $\sig_i$ as a counterclockwise exchange of the strands
in positions $i$ and $i+1$, with inverses of the generators
corresponding to clockwise twists. For instance, the braids
$\sig_1,\sig_2^{-1}\in B_3$ are given by
\[
\sig_1 = \braid{.5}{\r\s} \quad\quad\text{ and }\quad\quad
\sig_2^{-1} = \braid{.5}{\s\l}\,.
\]
The group product then corresponds to vertical concatenation of
braids (up to isotopy equivalence).  For instance,  
\[
\sig_2^{-1}\sig_1 = \braid{1}{\r\s\N\s\l}\,.
\]
The relation (\ref{braid1}) is known as the Yang-Baxter equation,
which is diagrammatically expressed as
\[
\braid{1.5}{
\r\s\N
\s\r\N
\r\s
} \;\;=\;\; \braid{1.5}{
\s\r\N
\r\s\N
\s\r
}\,.
\]
For each $n$, $B_n$ is isomorphic to the subgroup of $B_{n+1}$ generated by $\{\sig_1,\dotsc,\sig_{n-1}\}$, consisting of braids in which the $n+1$'st strand does not participate.  We denote the associated inclusion map as $\iota\colon B_n \to B_{n+1}$, which acts by adding an extra strand to the right of any given braid.  
Given a braid $b\in B_n$, there are many possible ways of
\emph{closing} the braid $b$ to obtain a link. In the following we
will recall three possible ways, the trace, the plat and a combination of these we call a
generalized closure.

\subsection*{Specifying links with braids}
One way to turn a braid into a link takes, for each $1\leq i \leq n$, the $i$'th
strand at the top of the braid and glues it to the $i$'th strand at the bottom.
The resulting link $\h{b}$ is called the \emph{trace
closure} of $b$.  For instance, the trace closure of the braid
$b=\sigma_2 \sigma_3^{-1} \sigma_2$ is isotopic to the union of the
trivial knot and the Hopf link:
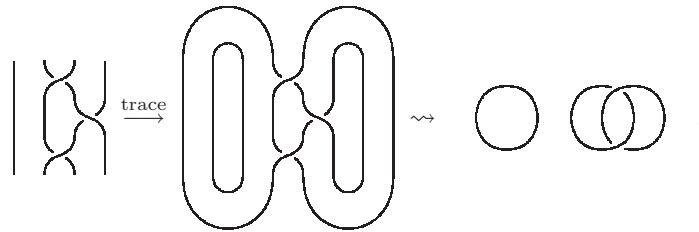
\[
\braid{1.5}{
\s\r\s\N
\s\s\l\N
\s\r\s
} \stackrel{\text{trace}}{\longrightarrow}
\braid{1.5}{
\caps{0}{3}\caps{1}{1}\caps{4}{3}\caps{5}{1}
\s\s\s\r\s\s\s\N
\s\s\s\s\l\s\s\N
\s\s\s\r\s\s\s\N
\cups{0}{3}\cups{1}{1}\cups{4}{3}\cups{5}{1}
} \rightsquigarrow
\xygraph{
!{0;/r1.0pc/:}
[u]
!{\hcap[2]}
!{\hcap[-2]}
[rrr]!{\vunder}
!{\vunder-}
[uur]!{\hcap[2]}
[l]!{\hcap[-2]}
}.
\]
Every link can be represented as the trace
closure of some braid.  Conventionally, one considers the trace
closure of a braid to yield an oriented link, since an orientation can
be unambiguously  defined to travel downward along the braid.  With this convention
every generator $\sigma_i$ gives rise to a positive crossing, while
its inverse $\sigma_i^{-1}$ to a negative crossing in the sense of Figure~\ref{figure:stuffJonesSkein}. 
A theorem of Markov demonstrates that, given braids $b$ and $b'$,
their trace closures $\h{b}$ and $\h{b'}$ are isotopic if and only if
$b$ and $b'$ are related by a finite sequence of \emph{Markov moves}.  We review these moves in Section~\ref{section:repformulae} as they are relevant for obtaining a representation-theoretic formulae for the one-variable HOMFLYPT and Jones polynomials of trace closures of braids. 

Another way of turning a braid on an even number $2p$ of strands into a
link is known as the \emph{plat closure}.  Such a procedure starts
with a braid $b\in B_{2p}$ and connects adjacent pairs of
strands at the top and at the bottom.  For example, the plat closure
of $b = \sigma_2\sig_3^{-1}\sig_2\in B_4$ is isotopic to the trefoil
knot, as
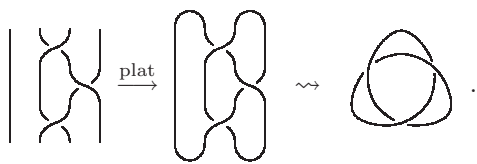
\[
\braid{1.5}{
\s\r\s\N
\s\s\l\N
\s\r\s
} \stackrel{\text{plat}}{\longrightarrow}
\braid{1.5}{
\caps{0}{1}\caps{2}{1}\s\r\s\N
\s\s\l\N
\s\r\s\N
\cups{0}{1}\cups{2}{1}
} \,\,\,\rightsquigarrow\,
\xygraph{
!{0;/r1.0pc/:}
!P3"a"{~>{}}
!P9"b"{~:{(1.3288,0):}~>{}}
!P3"c":{~:{(2.5,0):}~>{}}
!{\vunder~{"b2"}{"b1"}{"a1"}{"a3"}}
!{\vcap~{"c1"}{"c1"}{"b4"}{"b2"}}
!{\vunder~{"b5"}{"b4"}{"a2"}{"a1"}}
!{\vcap~{"c2"}{"c2"}{"b7"}{"b5"}}
!{\vunder~{"b8"}{"b7"}{"a3"}{"a2"}}
!{\vcap~{"c3"}{"c3"}{"b1"}{"b8"}}
}.
\]

Given a braid $b\in B_n$, we define a \emph{generalized closure} of $b$ to be
specified by the following data: two braids $x,y\in B_n$ and
nonnegative integers $p$ and $r$ which satisfy $2p+r = n$.  The
generalized closure $\chi_{x,y}^{p,r}(b)$ of $b$ is obtained by first
forming the braid $xby$, performing the plat closure on the $2p$
leftmost strands at the top and bottom, then connecting the remaining
$r$ strands at the top with the corresponding $r$ strands at the
bottom as in the trace closure. 

Unlike with the trace closure, there is no unambiguous way to assign an orientation to the plat or generalized closure of a braid (unless $n = t$).  While this poses a potential problem for defining the Jones and HOMFLYPT polynomials of such closures of braids, it is known \cite{lm87} (and easily proved using the Kauffmann bracket \cite{kauffman1} formula) that if $\overrightarrow{L}$ and $\overrightarrow{L'}$ are two oriented links that are isotopic as unoriented links, their Jones polynomials satisfy $J_{\overrightarrow{L}}(q) = q^m J_{\overrightarrow{L'}}(q)$ for some easily computable $m\in \bZ$.  This implies the absolute value of the Jones polynomial at roots of unity is independent of the orientation of a link, and is thus an invariant of unoriented links.  In Section~\ref{section:repformulae}, we provide a representation-theoretic formula for the absolute value of the Jones polynomial of any generalized closure of a braid at primitive roots of unity, while in Section~\ref{section:algorithms}, we present a quantum algorithm which approximates these invariants of unoriented links.

\subsection*{Additive approximations} 

The notion of approximation which we will require was recently formalized in \cite{bflw}; we review that material here.  Let $\CX$ be a set of problem instances, and suppose we are given a nonnegative function $f\colon \CX\to \bR$ which is potentially difficult to evaluate exactly.   The main idea of \cite{bflw} is to approximate the function $f$ with respect to some
positive normalization function $g\colon \CX\to \bR$.  Departing slightly from the definition given in \cite{bflw}, an \emph{additive approximation} for the normalized function $f/g$ associates a random variable $Z(x)$ to any  problem instance $x\in \CX$ and $\delta > 0$  satisfying
\[\Pr\left\{\left|\frac{f(x)}{g(x)} - Z(x)\right| \leq \delta\right\} \geq 3/4.\]
In addition, it is required that such an approximation be achieved in time which is polynomial in the size of the problem instance and in $1/\delta$. 

In Section~\ref{section:repformulae} we provide tight upper bounds on the evaluations of the invariants we wish to approximate.  We will see in (\ref{eqn:HOMFLYPTbound}) that if $b\in B_n$, the one-variable HOMFLYPT of the trace closure obeys
$\big|H^{(k)}_{\widehat{b}}(e^{2\pi i /\ell})\big| \leq [k]_\ell^{n-1}$.  Similarly, we show in (\ref{maxjones}) that if $n = 2p + t$ and $x,y\in B_n$, the Jones of the generalized closure satisfies 
$\big|J(\xi^{p,r}_{x,y}(b),e^{2\pi i /\ell})\big| \leq [2]_\ell^{p + t - 1}$.  These bounds are achieved by closing the identity braid for the one-variable HOMFLYPT of the trace closure, or otherwise the braid $x^{-1}y^{-1}$ for the Jones of the generalized closure.  
We use these normalizations for our additive approximations, so that the absolute values of the resulting normalized polynomials always lie between 0 and 1, regardless of the size of the braid, number of strands, or particular primitive root of unity. 
We therefore consider the following two problems:
\begin{problem}[\textbf{Approximate HOMFLYPT Trace Closure}]
\label{problem:approximatehomflypt}
Given is a braid $b\in B_n$ of length $m$, $\delta > 0$ and positive integers $k,\ell$ satisfying $2 \leq k < \ell$, where $k$ is a constant. The task is to sample from a random variable $Z\in \bC$, $|Z| \leq 1$, which is an additive approximation of the one-variable HOMFLYPT polynomial of the trace closure, evaluated at
$e^{2\pi i/\ell}$, in the sense that
\[\Pr\left\{\left|
\frac{1}{[k]_\ell^{n-1}}H^{(k)}_{\widehat{b}}(e^{2\pi i/\ell}) - Z \right| \leq \delta \right\} \geq 3/4.\]
\end{problem}

\begin{problem}[\textbf{Approximate Jones Closure}]
\label{problem:approximatejonesclosure}
Given is a braid $b\in B_n$ of length $m$, two braids $x,y\in B_n$ of length $O\big(\poly(m)\big)$, positive integers $\ell,p$ and $r$ which satisfy $2p+r=n$, and $\delta > 0$. The task is to
sample from a random variable $0\leq Z \leq 1$ which is an additive approximation of the absolute value of the Jones polynomial of the generalized closure, evaluated at
$e^{2\pi i/\ell}$, in the sense that
\[\Pr\left\{\left|
\frac{1}{[2]_\ell^{p+r-1}}\left|J\big(\chi_{x,y}^{p,r}(b),e^{2\pi i/\ell})\right| - Z \right| \leq \delta \right\} \geq 3/4.\]
\end{problem}
In Section~\ref{section:algorithms}, we prove:
\begin{theo} \label{thm:runningtime}
Each of the above two problems can be solved in $O(\poly(m,1/\delta))$ time on a quantum computer.
\end{theo}


\Section{Unitary Jones-Wenzl representations of the braid group} \label{section:reps}
Here we summarize the representation theory of the braid group that we will need, following \cite{wenzl}.  The relevant representations are inherited from representations of a certain family of Hecke algebras $H_n(q)$ which generalize the group algebra $\bC S_n$ of the symmetric group $S_n$.  In Section~\ref{section:repformulae}, we will show how these representations give rise to useful formulae for evaluations of the Jones and one-variable HOMFLYPT polynomials at roots of unity.
\subsection*{The Hecke algebra}
It is well-known that the group algebra $\bC S_n$ of the symmetric group $S_n$ on $\{1,2,\dotsc,n\}$ objects has a presentation in terms of generators $\{1,s_1,s_2,\dotsc,s_{n-1}\}$, where $s_i$ is the involution which swaps $i$ and $i+1$.  These satisfy the relations
\begin{eqnarray}
s_i^2 &=& 1  \label{s1}\\
s_is_{j}s_i &=& s_{j}s_is_{j}, \,\,\, |i-j| = 1 \label{s2}\\
s_i s_j &=& s_js_i, \,\,\,\,\,\,\,\, |i-j|>1. \label{s3}
\end{eqnarray}
For every $q\in \bC^{\times}$, the \emph{Hecke algebra $H_n(q)$ of type} $A_{n-1}$ is defined to be the algebra over $\bC$ generated by 1 and $\{g_1,\dotsc,g_{n-1}\}$ satisfying the relations
\begin{eqnarray}
g_i^2 &=& g_i(q-1) + q \hspace{.2in} \label{g1} \\
g_i g_{j} g_i &=& g_{j}g_ig_{j} \hspace{.586in} |i-j|=1 \label{g2} \\
g_ig_j &=& g_j g_i \hspace{.7in}  |i-j| > 1. \label{g3}
\end{eqnarray}
Note that this is a deformation of the group algebra $\bC S_n$ of the symmetric group $S_n$, which is obtained when $q=1$.
While it is known that $H_n(q)\isom\bC S_n$ whenever $q$ is not a root of unity, we will rather be interested in the cases when $q$ is a primitive root of unity, where the representation theory is slightly more subtle than that of $\bC S_n$.
One may represent the braid group $B_n$ inside the Hecke algebra $H_n(q)$ by simply mapping each generator $\sig_i$ of $B_n$ to the generator $g_i$ of $H_n(q)$.  Indeed, the representations of $B_n$ we will use in this paper are induced by representations of $H_n(q)$ on suitable finite-dimensional Hilbert spaces.
As with the braid groups, for each $n$, $H_n(q)$ is isomorphic to the subalgebra of $H_{n+1}(q)$ generated by $1$ and $\{g_1,\dotsc,g_{n-1}\}$.  We use the same notation for the inclusion maps as with the braid groups, writing $\iota\colon H_n(q) \to H_{n+1}(q)$.  One important distinction is that, writing $1_n$ for the identity in $H_n(q)$, we have $\iota(1_n) = 1_{n+1}$.  We will therefore often omit the subscript for the identity element of each $H_n(q)$ whenever there is no cause for confusion. 

\subsection*{Young diagrams and tableaux}
As with $S_n$, the unitary irreps of $B_n$ are labeled by
\emph{partitions} $\lambda = [\lambda_1,\lambda_2,\dotsc,\lambda_n]$
satisfying $\lambda_1\geq\lambda_2 \geq \cdots \geq \lambda_n \geq 0$ and $\sum_i \lambda_i  = n$. Throughout, we consider all partitions obtained by adding or deleting trailing zeros to be equivalent.  We will identify any such partition with its \emph{Young diagram} as pictured in Figure~\ref{young} for the partition $\lambda = [5,3,2]$.  Note that the diagram has one row for each \emph{part} of the partition, and that the $i$'th row contains $\lambda_i$ boxes.
\begin{figure} 
\[\yng(5,3,2)\]
\fcaption{\label{young}Young diagram for the partition $\lambda = [5,3,2]$.}
\end{figure}
We allow for the empty partition and diagram, which we denote $\emptyset$.
Let $\Lambda_n$ be the collection of $n$-box Young diagrams.
Given two diagrams $\lambda\in \Lambda_n$ and $\mu\in \Lambda_m$, their \emph{sum} $\lambda+\mu\in \Lambda_{n+m}$ is always well-defined, having parts equal to $(\lambda + \mu)_i = \lambda_i + \mu_i$.  When it is well-defined as a partition, we will similarly define the \emph{difference} $\lambda - \mu\in\Lambda_{n-m}$ to have parts $(\lambda - \mu)_i = \lambda_i - \mu_i$.

Define the sets of diagrams $\Lambda_{m,n} \equiv \biguplus_{i=m}^n \Lambda_i$ and abbreviate $\Lambda \equiv \Lambda_{0,\infty}$.
We may consider elements of $\Lambda$ to be the nodes of an infinite directed graph, which we will call the \emph{Young graph}, which contains an edge from a diagram $\lambda'$ to a diagram $\lambda$ whenever a single box can be added to $\lambda'$ to obtain $\lambda$.  We remark that we depart slightly from common terminology, where this graph, with each arrow reversed, is called a \emph{Bratteli diagram}.
With a slight abuse of notation, we write $\Lambda$ for both the set of diagrams and for the corresponding Young graph, making a similar identification between subsets of $\Lambda$ such as $\Lambda_{0,n}$ and the associated subgraph.
In Figure~\ref{younggraph}, we give a picture of $\Lambda_{0,4}$.
\begin{figure} 
\begin{tiny}
\[\xymatrix@R=.5cm@C=.3cm{
& & & & \emptyset \ar[d]\\
& & & & {\yng(1)} \ar[dr] \ar[dl]\\
& & & {\yng(2)}\ar[dl]\ar[dr] & &  {\yng(1,1)}\ar[dl]\ar[dr] \\
& & {\yng(3)}\ar[dll]\ar[d] & & {\yng(2,1)}\ar[dll]\ar[d]\ar[drr] & & {\yng(1,1,1)}\ar[d]\ar[drr] \\
{\yng(4)} & & {\yng(3,1)} & & {\yng(2,2)} & &  {\yng(2,1,1)} & & {\yng(1,1,1,1)}
}\]
\end{tiny}
\fcaption{\label{younggraph}The Young graph $\Lambda_{0,4}$.} 
\end{figure}

By a \emph{numbering} of an $n$-box diagram, we mean any assignment of the numbers $\{1,2,\dotsc, n\}$ to the boxes of the diagram, where each number appears only once.
For a given diagram $\lambda \in \Lambda_n$, we write $T_\lambda$ for the set of \emph{standard tableaux} with shape $\lambda$, corresponding to numberings which are strictly increasing along each row and column.  We often call a standard tableau just a \emph{tableau}.  We denote the set of all $n$-box tableaux as $T_n = \biguplus_{\lambda\in \Lambda_n} T_\lambda$ and the set of all tableaux as $T = \biguplus_{n=1}^{\infty} T_n$.  We identify members of $T_n$ with length-$n$ paths starting at $\emptyset$ in $\Lambda_{0,n}$, pictured in Figure~\ref{younggraph}, with elements of $T_\lambda$ correspond to paths which end at the diagram $\lambda$.

In this paper, we shall rather require certain restricted classes of Young diagrams and tableaux.  To each pair $(k,\ell)$ of integers satisfying $\ell > k > 0$ corresponds a class of irreducible unitary representations of $B_n$.  These representations are labelled by the $n$-box \emph{$(k,\ell)$-Young diagrams}, defined as
\[\Lambda_n^{(k,\ell)} = \{\lambda\in \Lambda_n : \lambda_{k+1} = 0, \lambda_1 - \lambda_k\leq \ell-k\}.\]
Note that these are the diagrams with at most $k$ rows for which the difference between the numbers of boxes in the first and $k$'th rows is at most $\ell-k$.  We will often just say that such a $\lambda$ is an $(k,\ell)$-diagram.  For a given diagram $\lambda$, we refer to $\lambda_1 - \lambda_k$ as the \emph{level} of $\lambda$, so that $\ell-k$ is the maximum level of all diagrams in $\Lambda_n^{(k,\ell)}$.
As before, we make the similar abbreviations $\Lambda^{(k,\ell)}_{m,n}$ and $\Lambda^{(k,\ell)}$. We also identify any of these sets of diagrams with the appropriate subgraph of $\Lambda$.  We refer to such a graph as a $(k,\ell)$-\emph{Young graph}.
In Figure~\ref{klyounggraph}, we show the graph $\Lambda^{(2,5)}_{0,4}$.
\begin{figure}
\begin{tiny}
\[\xymatrix@R=.5cm@C=.3cm{
& & & & \emptyset \ar[d]\\
& & & & {\yng(1)} \ar[dr] \ar[dl]\\
& & & {\yng(2)} \ar[dr] \ar[dl]& &  {\yng(1,1)}\ar[dl] \\
& & {\yng(3)} \ar[d]& & {\yng(2,1)}\ar[d]\ar[dll]  \\
& & {\yng(3,1)}  & & {\yng(2,2)}
}\]
\end{tiny}

\fcaption{\label{klyounggraph}The Young graph $\Lambda_{0,4}^{(2,5)}$.  Comparison with Figure~\ref{younggraph} reveals that this is the subgraph of $\Lambda_{0,4}$ obtained by removing vertices whose diagrams either have more than two rows, or level greater than $3=5-2$.}
\end{figure}
For a given $\lambda\in \Lambda^{(k,\ell)}_n$, define the \emph{$(k,\ell)$-tableaux} $T_\lambda^{(k,\ell)}\subset T_\lambda$ to be those standard tableaux of shape $\lambda$ for which successively deleting the largest numbered boxes yields, at each step, another $(k,\ell)$-diagram. These tableaux can be identified with paths in $\Lambda_{0,n}^{(k,\ell)}$ going from $\emptyset$ to $\lambda$.  We also define $T_n^{(k,\ell)}$ and $T^{(k,\ell)}$ in a similar manner as above.
In the Section~\ref{section:repformulae}, we will see that the $(2,\ell)$-diagrams and tableaux are what is needed to give representation-theoretic expressions for the Jones polynomials $J(e^{2\pi i/\ell})$ of closures of braids, while the $(k,\ell)$-diagrams are relevant for the one-variable HOMFLYPT polynomials $H^{(k)}(e^{2\pi i/\ell})$.

\subsection*{Unitary Jones-Wenzl representations of $B_n$}

The representations of $B_n$ we will introduce are parameterized by integers $\ell > k > 0$, where $\ell$ specifies a primitive $\ell$'th root of unity which we will denote $q = e^{2\pi i/\ell}$. The irreducible components of the representations we will describe reduce to the usual irreps of $S_n$ when $k=n$ and $\ell \to \infty$, so that $q\to 1$.  For the rest of this section, we fix $\ell > k > 0$ and set $q = e^{2\pi i/\ell}$, describing a corresponding representation of $H_n(q)$ which induces a unitary representation of $B_n$.
This representation is more easily described by a change of variables which leads to an equivalent presentation of $H_n(q)$.
Rewriting the quadratic relation (\ref{g1}) as $(g_i + 1)(g_i - q) = 0$, we may define idempotents $e_i = (q-g_i)/(1+q)$ (meaning that $e_i^2 = e_i$) for which
\begin{eqnarray} \label{etog}
g_i = q(1-e_i) - e_i = q - (1+q) e_i.
\end{eqnarray}
Using these idempotents, the relations (\ref{g1}--\ref{g3}) may be expressed as
\begin{eqnarray}
e_i^2 &=& e_i \hspace{.2in}  \label{e1} \\
e_i e_j e_i -  \tau e_i &=& e_{j}e_ie_{j} -  \tau e_j, \hspace{.52in} |i-j| = 1 \label{e2}\\
e_ie_j &=& e_j e_i, \hspace{1in} |i-j| > 1, \label{e3}
\end{eqnarray}
where we follow the usual convention in setting $\tau\equiv [2]_\ell^{-2}$.
For each integer $d$, define
\begin{equation}
a_{\ell}(d) = \frac{[d+1]_\ell}{[2]_\ell[d]_\ell},
\end{equation}
noting that $a_\ell(d) + a_\ell(-d) = 1=a_\ell(1)$.
Given any $n$-box tableau $t$ and an integer $1\leq i < n$, we respectively write $c_i(t)$ and $r_i(t)$ for the \emph{column} and \emph{row} which contain the number $i$, defining
\begin{eqnarray}
d_i(t) \equiv c_i(t) - c_{i+1}(t) - (r_{i}(t) - r_{i+1}(t)). \label{dit}
\end{eqnarray}
Note that $d_{i}(t)$ gives the total number of leftward and downward steps it takes to ``walk" from $i$ to $i+1$ on $t$, where moving right or up counts negatively.
Given a generator $s_i$ of $S_n$ 
and a $(k,\ell)$-tableaux $T_n^{(k,\ell)}$, let $t'$ be the numbering obtained by 
swapping the numbers $i$ and $i+1$ in $t\in T_n^{(k,\ell)}$.  If $t'$ is also a $(k,\ell)$-tableau, define $s_i(t) = t'$ and $s_i(t') = t$;
otherwise set $s_i(t) = t$.  In order that $s_i(t)\neq t$, it is necessary and sufficient that $i$ and $i+1$ \emph{not} be in the same row or column of $t$.  In such a case, we may think of it as a deformation of the path in $\Lambda^{(k,\ell)}_{0,n}$ corresponding to $t$ which changes only the $i$-box diagram passed in the original path.  
Note that each generator $s_i$ induces a partition of $T^{(k,\ell)}_n$ into sets of numberings (or paths) of size at most two. 

 We may now define a representation $\pi_n^{(k,\ell)}$ of $H_n(q)$ on a vector space  $V^{(k,\ell)}_n$ which has an orthonormal basis labeled by $n$-box $(k,\ell)$ tableaux, i.e.\
\[V^{(k,\ell)}_n = \text{span}\big\{\ket{t} : t \in T^{(k,\ell)}_n\big\}.\]
For each $i$, $\pi^{(k,\ell)}_n(e_i)$ is a projection which is block diagonal in the $T_n^{(k,\ell)}$ basis with blocks of size either one or two.
The size two blocks correspond to distinct pairs of tableaux $t$ and $s_i(t)$ which are contained in $T_n^{(k,\ell)}$. Letting $V_{i,t}=\text{span}\big\{\ket{t},\ket{s_i(t)}\}$,    the restriction of $\pi_n^{(k,\ell)}(e_i)$ to $V_{i,t}$ is the rank one projection onto $\sqrt{a_\ell\big(d_i(t)\big)}\ket{t} + \sqrt{a_\ell\big(\!\!-\!d_i(t)\big)}\ket{s_{i}(t)}$, written
\begin{equation}
\pi_n^{(k,\ell)}(e_i)\big|_{V_{i,t}} = \pmat{a_\ell\big(d_i(t)\big) & \sqrt{a_\ell\big(d_i(t)\big)a_\ell\big(\!\!-\!d_i(t)\big)} \\
\sqrt{a_\ell\big(d_i(t)\big)a_\ell\big(\!\!-\!d_i(t)\big)} & a_\ell\big(\!\!-\!d_i(t)\big)}.
\end{equation}
For the remaining $t\in T^{(k,\ell)}_n$, set $\pi_n^{(k,\ell)}(e_i)\ket{t} = \ket{t}$ if $i$ and $i+1$ are in the same column of $t$; otherwise set $\pi_n^{(k,\ell)}(e_i)\ket{t} = 0$.  This defines the action of $\pi_n^{(k,\ell)}(e_i)$ on all of $V^{(k,\ell)}_n$.
Using (\ref{etog}), this representation maps each $g_i$ to the unitary matrix
\begin{eqnarray} \label{pikl}
\pi_n^{(k,\ell)}(g_i)=
q1_{V^{(k,\ell)}_n} -(1+q) \pi_n^{(k,\ell)}(e_i).
\end{eqnarray}
Abbreviating $d = d_i(t)$, we may use (\ref{eqn:qintidentity}) to express the restriction to each $V_{i,t}$ as
\begin{eqnarray} 
\pi_n^{(k,\ell)}(g_i)\big|_{V_{i,t}} =  
-q^{1/2}\pmat{\frac{q^{d/2}}{[d]_\ell} & \sqrt{1 - \frac{1}{[d]_\ell^{2}}} \\ 
\sqrt{1 - \frac{1}{[d]_\ell^{2}}} & -\frac{q^{d/2}}{[d]_\ell}}.
\label{eqn:theunitary}
\end{eqnarray}
If $i$ and $i+1$ are in the same column of $t$, this gives $\pi_n^{(k,\ell)}(g_i) \ket{t}=-\ket{t}$, while if they are in the same row we have $\pi_n^{(k,\ell)}(g_i) \ket{t}=q\ket{t}$.
Because permutations cannot change the shapes of tableaux, it follows that this representation is reducible.  The irreducible components are labeled by the $(k,\ell)$-diagrams, so that
\[\pi_n^{(k,\ell)} = \bigoplus_{\lambda\in \Lambda^{(k,\ell)}_n} \pi_\lambda^{(k,\ell)},\]
where each $\pi_\lambda^{(k,\ell)}$ acts nondegenerately only on the subspace $V^{(k,\ell)}_\lambda = \big\{\ket{t} : t\in T_\lambda^{(k,\ell)}\}$, and
\[V^{(k,\ell)}_n = \bigoplus_{\lambda\in \Lambda_n^{(k,\ell)}}V^{(k,\ell)}_\lambda.\]

%
%
\section{Efficient quantum circuits for unitary Jones-Wenzl representations}
\label{section:localqubitmodel}
We will now show, for any integers $1 < k < \ell$, how to implement
the corresponding Jones-Wenzl representations $\pi_n^{(k,\ell)}$ defined in
(\ref{pikl}) on a quantum computer. The basic idea is to embed these representations into a tensor product space (c.f.\ Section~5.2 of \cite{sb}).  
Our circuits resemble those of \cite{ajl}, where the $k=2$ case is treated, although we require the combinatorial tools of Young diagrams and tableaux introduced in the previous section for higher values of $k$.  These circuits allow us to give a new quantum algorithm in Section~\ref{section:algorithms} for  approximating the family of one-variable HOMFLYPT polynomials $H^{(k)}$ of the trace closures of braids.   We begin with some background on quantum circuits.  

The state space of a quantum computer
has the structure of a Hilbert space $\CH$.  Denoting by $U(\CH)$ the
group of unitary matrices acting on $\CH$, a \emph{quantum gate} is
any $U\in U(\CH)$.  
By the \emph{standard quantum circuit model}, we shall mean quantum computers
whose state space decomposes into a finite number of localized
two-dimensional subsystems, or qubits, so that $\CH = (\bC^2)^{\otimes
n}$.  We fix a preferred orthonormal basis for the Hilbert space of each qubit, which we call the \emph{computational basis}:
\[\ket{0} \equiv \pmat{1\\0}\,\,\,\,\,\,\,\,\,\,\ket{1} \equiv \pmat{0\\1}.\]
The state space of $\CH$ is thus spanned by tensor products of these basis vectors, which we write as $\big\{\ket{x^n} : x^n \in \{0,1\}^n\big\}$, where 
\[\ket{x^n} \equiv \ket{x_1}\ket{x_2}\cdots \ket{x_n}\equiv \ket{x_1}\otimes \ket{x_2}\otimes \cdots \otimes \ket{x_n}.\]
We will abbreviate the set of all gates on $n$ qubits as $U(2^n)
\equiv U\big((\bC^2)^{\otimes n}\big)$.  We assume that the qubits are arranged on a line so that each qubit has at most two neighbors.  In this case, it is well-known that an arbitrary unitary $U\in U(2^n)$ can be performed on the qubits by a sequence of (possibly exponentially many in $n$) \emph{local unitaries} from $U(4)$ which act only on adjacent pairs of qubits.
By a \emph{length-$m$ quantum circuit}, we will mean a sequence of $m$ local two-qubit gates.  When appropriate, we identify a quantum circuit with its overall unitary transformation $U = U_mU_{m-1}\cdots U_1$, where each $U_i$ is a two-qubit unitary. 
Below, we will work with quantum circuits whose local systems (registers) may have dimension $k > 2$.   Such systems can be implemented using $O(\log k)$ qubits, on which any unitary can be performed using only $O(k)$ two-qubit gates.

We now proceed  by describing an encoding of the basis states of the
reducible representation $V^{(k,\ell)}$ into $O\big(n\log(k)\big)$ qubits.  Then we show how, using $O\big(n\log(n)\big)$ ancillary qubits, one can implement
the image $\pi_n^{(k,\ell)}(\sig_i)$ of each braid group generator
$\sig_i$ with a quantum circuit of length $O\big(\poly(n,k)\big)$.
Given a tableau $t\in T^{(k,\ell)}_n$ and an integer $1\leq i\leq n$,
recall that we write $r_i(t)$ and $c_i(t)$ for the row and column
which respectively contain the number $i$.  Notice that these numbers satisfy $1\leq r_i(t) \leq k$ and $1\leq c_i(t) \leq n$ for each
$i$. We abbreviate $r(t) =
r_1(t)r_2(t)\cdots r_n(t)\in \{1,\dotsc,k\}^n$ and define $c(t)\in \{1,\dotsc,n\}^n$ similarly.
Our first observation is that $r(t)$ uniquely specifies $t$. We may
thus assign to each $t$ a computational basis state of $n$ systems with
Hilbert spaces $\CR_1,\CR_2,\dotsc,\CR_n$, where each Hilbert space
$\CR_i\isom \bC^k$ is spanned by computational basis states
$\{\ket{1}^{\CR_i},\dotsc,\ket{k}^{\CR_i}\}$.  Together, the systems have a
combined Hilbert space which we write $\CR^n = (\bC^k)^{\otimes n}$.
This yields an embedding $\ket{t}
\mapsto \ket{r(t)}^{\CR^n}$ of $V_n^{(k,\ell)}$ into $\CR^n$ whose
image we call the \emph{computational subspace}.  

 We now introduce, for each $1\leq i\leq n-1$, a register with Hilbert space $\CD_i$ which will store the value $d_i(t)$ defined in (\ref{dit}) for each tableau.  Because $|d_i(t)| \leq n$, we let $\CD_i$ be spanned by basis states $\{\ket{\!-\!n}^{\CD_i}, \ket{\!-\!n\!+\!1}^{\CD_i}, \dotsc,\ket{n}^{\CD_i}\}$.
We may now define an isometry $W\colon \CR^n \to \CR^n\CD^{n-1}$ which coherently computes these values for a given encoding of a tableau as 
\begin{equation}
W\ket{r(t)}^{\CR^n} = \ket{r(t)}^{\CR^n}\ket{d(t)}^{\CD^{n-1}}. \label{prepareD}
\end{equation}
Below, we will show that this isometry can be implemented efficiently.
Now we follow  (\ref{eqn:theunitary}) to define the unitaries which perform the braiding, while taking care to appropriately update the $\CD^{n-1}$ registers.  
For each $2\leq i \leq n-2$, define the unitary $U'_i$ which acts on $\CR_i\CR_{i+1}\CD_{i-1}\CD_i\CD_{i+1}$ as
\begin{eqnarray*}
U'_i\ket{r_i,r_{i+1},d_{i-1},d_i,d_{i+1}} =
\al_{d_i} \ket{r_i,r_{i+1},d_{i-1},d_i,d_{i+1}}
+ \beta_{d_i}\ket{r_{i+1},r_i,d_{i-1}\!+\!d_i,\!-d_i,d_{i+1}\!+\!d_i} .
\end{eqnarray*}
where $\al_d = -\frac{q^{-\frac{d-1}{2}}}{[d]_\ell}$ and $\beta_d =  -q^{1/2}\sqrt{1 - \frac{1}{[d]_\ell^2}}$.
Since $d_1$ can only equal $\pm 1$, we also set  
\[U'_1\ket{d_1}^{\CD_1} = 
\begin{cases} 
-\ket{d_1}^{\CD_1} & \text{ if } d_1 = 1 \\
q\ket{d_1}^{\CD_1} & \text{ if } d_1= -1.
\end{cases}\]
Additionally, we define 
$U_{n-1}'$ to act on $\CR_{n-1}\CR_{n}\CD_{n-2}\CD_{n-1}$ via 
\[U_{n-1}'\ket{r_{n-1},r_n,d_{n-2},d_{n-1}} = 
\al_{d_{n-1}} \ket{r_{n-1},r_n,d_{n-2},d_{n-1}} + \beta_{d_{n-1}} \ket{r_{n-1},r_n,d_{n-2}\!+\!d_{n-1},\!-d_{n-1}}.\]
Each of these unitaries can be implemented using $O\big(\poly(n,k)\big)$ two-qubit gates.  
Define the unitaries $U_i = W^{-1}U'_iW$ for $1\leq i < n$.
By construction, the map $\sig_i \mapsto U_i$ defines a (degenerate) unitary representation $\rho^{(k,\ell)}_n$ of $B_n$ which is isomorphic to $\pi_n^{(k,\ell)}$.
Suppose now that we are given a braid $b=\sig_{i_m}^{x_m}\cdots \sig_{i_2}^{x_2}\sig_{i_1}^{x_1}\in B_n$.
The corresponding unitary
\begin{eqnarray*}
\sig_{i_m}^{x_m}\cdots \sig_{i_2}^{x_2}\sig_{i_1}^{x_1}&\mapsto& U_{i_m}^{x_m}\cdots U^{x_2}_{i_2}U^{x_1}_{i_1} \\
&=& W^{-1}U_{i_m}'^{x_m}\cdots U_{i_2}'^{x_2}U_{i_1}'^{x_1}W
\end{eqnarray*}
therefore evolves the computational subspace of $\CR^n$ according to the following commutative diagram
\[\xymatrix@C=2cm@R=.5cm{
V^{(k,\ell)}_n \ar[r]^{\pi_n^{(k,n)}(b)} \ar[d] & V^{(k,\ell)}_n \\
\CR^n \ar[r]_{\rho^{(k,\ell)}_n(b)} & \CR^n. \ar[u]
}\]

Now we show that the isometry $W$ in (\ref{prepareD}) can be efficiently implemented.  For this, we introduce $n$ systems with joint Hilbert space $\CC^n$, where each
$\CC_i$ is spanned by computational basis states
$\{\ket{0}^{\CC_i},\dotsc,\ket{n}^{\CC_i}\}$.  The purpose of these
systems is to store the column numbers of each $i$.  In addition, we
introduce $k$ registers $\CL_1,\dotsc,\CL_k$ with bases
$\{\ket{0},\dotsc,\ket{n}\}$ which will store the \emph{shape} of the
tableau.  Given a qubit encoding $\ket{r(t)}^{\CR^n}$ of a tableau
$t$, we define the isometry $W'\colon \CR^n \to \CR^n\CC^n
\CL^k$ as
\[W'\ket{r(t)}^{\CR^n} = \ket{r(t)}^{\CR^n}\ket{c(t)}^{\CC^n}\ket{\lambda(t)}^{\CL^k},\]
where $\lambda(t)$ is the vector of row lengths of $t$.
This isometry can be efficiently implemented by first preparing the additional systems in the state $\ket{0\cdots 0}^{\CC^n}\ket{0\cdots 0}^{\CL^k}$, after which the composition of local gates  $W' = W'_n \cdots W'_2 W'_1$ is applied, where each $W'_i$ acts on $\CR_i\CC_i\CL^k$ as
\[W'_i\ket{r}^{\CR_i}\ket{0}^{\CC_i}\ket{\lambda}^{\CL^k} =
\ket{r}^{\CR_i}\ket{\lambda_r\!+\!1}^{\CC_i}\ket{\lambda'}^{\CL^k}
\]
where $\lambda'$ is $\lambda$ with an additional box in row $r$.
For each $1\leq i < n$, we then define local unitaries $W_i$ which act as $\CR_i\CR_{i+1}\CC_i\CC_{i+1}\CD_i$ as
\[W_i
\ket{r}^{\CR_i}\ket{r'}^{\CR_{i+1}}\ket{c}^{\CC_i}\ket{c'}^{\CC_{i+1}}\ket{0}^{\CD_i} =
\ket{r}^{\CR_i}\ket{r'}^{\CR_{i+1}}\ket{c}^{\CC_i}\ket{c'}^{\CC_{i+1}}\ket{c - c' - (r - r')}^{\CD_i}.\]
We may thus write $W = W'^{-1}W_{n-1}\cdots W_2 W_1  W'$, where we uncompute the $\CC^n$ and $\CL^k$ registers.  Because each $W_i$ and $W'_i$ can be implemented with $O\big(\poly(n,k)\big)$, we find that $W$ can be implemented with $O\big(\poly(n,k)\big)$ local gates.

%
%
\section{Representation-theoretic formulae for the Jones and HOMFLYPT} \label{section:repformulae}
We will now illustrate how to use the unitary representations of Section~\ref{section:reps} to write expressions for the evaluations, at primitive roots of unity, of the Jones and one-variable HOMFLYPT polynomials of the trace closure of a braid.
We then review the Temperley-Lieb algebra, which we use to derive a formula for the Jones polynomial of our generalized closure of a braid.

\subsection*{Jones and HOMFLYPT polynomials of trace closures from the Markov trace}

Here, we review the theory of Markov traces on Hecke algebras and show how they yield formulae for the one-variable HOMFLYPT polynomials of the trace closure of a braid.  
Our presentation is primarily based on Section~6 of \cite{jones87} and Section~3 of \cite{wenzl}.  We urge the reader who seeks a deeper understanding to consult these references.  We also recommend \cite{bigelow}.  Throughout this section, we fix a primitive root of unity $q=e^{2\pi i/\ell}$ for some integer $\ell \geq 3$.

The theory of Markov traces is best understood in light of the so-called Markov moves on braids.  As mentioned earlier, the trace closures of two braids are isotopic iff the braids are connected by a finite sequence of Markov moves.  
Given any $b\in B_n$, a \emph{type I Markov move}
consists of conjugating $b$ by any braid $x\in B_n$, taking $b\mapsto
xbx^{-1}$.  A \emph{type II Markov move} adds another strand to $b$ which is
twisted either clockwise or counterclockwise with the $n$'th strand at
the bottom of $b$, acting as $b\mapsto \sig_n^{\pm} \iota(b)$.  It is clear
that the trace closures of braids related by the first Markov move are
isotopic, since the top braid $x$ is untied by the bottom braid
$x^{-1}$.  A picture more clearly illustrates the invariance of the
trace closure under the second Markov move, as shown in
Figure~\ref{figure:secondMarkovmove}.

%
%
\begin{figure}
\vspace{-.3in}
\[
\braid{2}{\caps{0}{3}\caps{1}{1};(1,2) \s\s\s; (1,1)\s\s\s; (2,0)\s\s;
(0,0) \r; (1,-1)\vcap- (0,-1)\vcap[-3];
(-2.2,0) *{} ="a"; (.2,0) *{}="b"; (.2,2) *{}="c"; (-2.2,2) *{}="d";
"a";"b" **\dir{-};
"b";"c" **\dir{-};
"c";"d" **\dir{-};
"d";"a" **\dir{-};
(-2,0) \s\s; (-2,3)\s\s; (-1,1) *{b};
(1.2,2)*{}="e";(1.2,-.8)="f"; (-2.2,-.8)="g";
"c";"e" **\dir{.};
"e";"f" **\dir{.};
"f";"g" **\dir{.};
"a";"g" **\dir{.};
} \,\,\, \rightsquigarrow\,\,\,
\braid{2}{\caps{0}{1}; (1,2)\s;(1,1)\s;(0,0)\vcap-;
\POS (-2.2,0) *{} ="a"; (.2,0) *{}="b"; (.2,2) *{}="c"; (-2.2,2) *{}="d";
"a";"b" **\dir{-};
"b";"c" **\dir{-};
"c";"d" **\dir{-};
"d";"a" **\dir{-};
(-2,0) \s\s; (-2,3)\s\s; (-1,1) *{b};
} \vspace{-.3in}
\]
\fcaption{\label{figure:secondMarkovmove}Trace closure after the second Markov move on some $b\in
B_3$. The dotted box on the left represents the braid $\sig_3\iota(b)\in B_4$.  We
have omitted the strands which close the two leftmost strands for
visibility.}
\end{figure}

These topological considerations are reflected algebraically as follows. 
It is a theorem of Ocneanu \cite{homfly} (see \cite{jones87} for a simple proof) that for every $1\leq k < \ell$
and for each $n$, there is a linear function $\tr_{k\ell} \colon H_n(q) \to \bC$ uniquely determined by
\begin{enumerate}
\item $\tr_{k\ell}(1_n) = 1$ for $1_n\in H_n(q)$
\item $\tr_{k\ell}(wv) = \tr_{k\ell} (vw)$ for $w,v\in H_n(q)$
\item $\tr_{k\ell}( e_n\iota(w)) =   \frac{[k-1]_\ell}{[k]_\ell [2]_\ell} \tr_{k\ell}(w)$ for all $w\in H_n(q)$
for all $n$\,.
\end{enumerate}
Using (\ref{etog}), a direct calculation shows that 3.\ implies that for $w\in H_n(q)$, 
\begin{equation}
\tr_{k\ell}(g_n\iota(w)) = \left(q - (1+q) \frac{[k-1]_\ell}{[k]_\ell [2]_\ell}\right)\tr_{k\ell}(w) = \frac{q^{\frac{k+1}{2}}}{[k]_\ell}\tr_{k\ell}(w).
\label{eqn:traceg}
\end{equation}
Such a function is known as a \emph{positive Markov trace} on the inductive limit $H_\infty(q)$ of the Hecke algebras $H_n(q)$. This trace has a representation-theoretic formula which we recall below in (\ref{weightedtrace}).  But first, we show how it can be used to express the Jones and one-variable HOMFLYPT polynomials.  One first extends this trace to $B_n$ via the  
identification
$\sig_i^{\pm} \mapsto g_i^{\pm}$.  Cyclicity of the trace implies that $\tr_{k\ell}$ is invariant under type I Markov moves.  However, it is not invariant under the second move because (\ref{eqn:traceg}) implies that multiplication of $w\in H_n(q)$ by either $g_n$ or $g_n^{-1}$ affects the trace nontrivially.  We will therefore construct, for each $n$, a normalized version $\overline{\tr}_{k\ell}\colon H_n(q) \to \bC$ of the Markov trace
which is also invariant under type II Markov moves, i.e.\ 
\begin{equation} \label{normalizedtrace}
\overline{\tr}_{k\ell}(\sigma^{\pm 1}_n \iota(b)) = \overline{\tr}_{k\ell}(b)
\end{equation}
for each $b\in B_n$.  For this, we define 
\begin{eqnarray}
\overline{\tr}_{k\ell}(b) 
&=& [k]_\ell^{n-1} q^{-\frac{k+1}{2}e(b)} \tr_{k\ell}(b). \label{eq:HOMFLYPT}
\end{eqnarray}
Here, $e(b)$ is the sum of the exponents of the
generators in the expression of $b$.
Now if $b \in B_n$ and $b' = \sig_n^{\pm 1}\iota(b)\in B_{n+1}$,  
\begin{eqnarray*}
\overline{\tr}_{k\ell}(b') &=&  [k]_\ell^{n} q^{-\frac{k+1}{2}e(b')} \tr_{k\ell}(b') \\
&=& [k]_\ell^{n} q^{-\frac{k+1}{2}\left(e(b)\pm 1\right)}\frac{q^{\pm\frac{k+1}{2}}}{[k]_\ell} \tr_{k\ell}(b) \\
&=& [k]_\ell^{n-1} q^{-\frac{k+1}{2}e(b)}\ \tr_{k\ell}(b) \\ 
&=& \overline{\tr}_{k\ell}(b)
\end{eqnarray*}
as required.  
To see that this normalized trace is equal to the one-variable HOMFLYPT of the trace closure of $b$, we follow \cite{jones87} by considering an arbitrary oriented link diagram $L_0$ described as the trace closure of some braid.  Choose a crossing in the braid arbitrarily -- we assume it to be positive without loss of generality.   By a type I Markov move, one can then express $L_0$ as the trace closure of a braid where the given positive crossing is at the bottom, i.e.\ $L_0 = \widehat{\sig_i b}$.  Inserting a positive or negative crossing adjacent to the one specified above results in the oriented links
$L_+ = \widehat{\sig_i^2 b}$ and $L_- = \widehat{b}$.  Using the formula (\ref{eq:HOMFLYPT}) to express the normalized trace in terms of the linear trace $\tr_{k\ell}$, and replacing $\sig_i^2$ using the quadratic relation (\ref{g1}), it is straightforward to check that 
\[q^{k/2}\overline{\tr}_{k\ell}(\sig_i^2 b) = q^{-k/2}\overline{\tr}_{k\ell}(b) + (q^{1/2} - q^{-1/2})\overline{\tr}_{k\ell}(\sig_i b).\]
Therefore, the normalized trace satisfies the definition (\ref{eqn:1varHOMFLYPTdefn})  of the one-variable HOMFLYPT polynomial from the skein relation (\ref{eqn:HOMFLYPTskein}) with $H^{(k)}_{\widehat{\sig_i b}}(q) = \overline{\tr}_{k\ell}(\sig_i b)$, so that 
\begin{equation}
H^{(k)}_{\widehat{b}}(q) =  [k]_\ell^{n-1} q^{-\frac{k+1}{2}e(b)} \tr_{k\ell}(b).
\label{eqn:HOMFLYPTtrace}
\end{equation}
For $w\in H_n(q)$, the Markov trace has the following representation-theoretic formula
\begin{equation} \label{weightedtrace}
\tr_{k\ell}(w) = \sum_{\lambda\in \Lambda_n^{(k,\ell)}} s^{(k,\ell)}_\lambda \Tr \pi^{(k,\ell)}_\lambda(w).
\end{equation}
Here, the \emph{Markov weights} $s^{(k,\ell)}_\lambda$ are Schur functions given explicitly by
\begin{equation}
s^{(k,\ell)}_\lambda = \frac{1}{[k]^n_\ell}\prod_{(i,j)\in \lambda}\frac{[j-i+k]_\ell}{ [h(i,j)]_\ell} 
\label{skl}
\end{equation}
for each $\lambda \in \Lambda_n^{(k,\ell)}$
and $h(i,j)$ denotes the \emph{hook length} of the box with row-column coordinates $(i,j)$ in $\lambda$.
For each $\lambda \in \Lambda_n^{(k,\ell)}$, if we write
$d_\lambda^{(k,\ell)}$ for the dimension of $V_{\lambda}^{(k,\ell)}$,
we may evaluate (\ref{weightedtrace}) at the identity $1_n\in H_n(q)$ to obtain the following formula relating the Markov weights and the dimensions of the irreps:
\begin{equation} \label{weightsdims1}
\sum_{\lambda\in \Lambda^{(k,\ell)}_n} s_\lambda^{(k,\ell)} d_\lambda^{(k,\ell)} = \tr_{k\ell}(1_n) = 1.
\end{equation}
We are now equipped to verify the bound on the HOMFLYPT given at the end of Section~\ref{section:links}.  Indeed, by unitarity we may combine (\ref{eqn:HOMFLYPTtrace}), (\ref{weightedtrace}) and (\ref{weightsdims1}) to obtain, for each $b\in B_n$,  
\begin{eqnarray}
\big|H^{(k)}_b(e^{2\pi i/\ell})\big| &=& 
[k]^{n-1}_\ell \Big|\sum_{\lambda\in \Lambda^{(k,\ell)}_n} s_\lambda^{(k,\ell)} \Tr\pi_\lambda^{(k,\ell)}(b)\Big| \nn \\
&\leq& [k]^{n-1}_\ell \sum_{\lambda\in \Lambda^{(k,\ell)}_n} s_\lambda^{(k,\ell)} d_\lambda^{(k,\ell)} \nn \\
&=&  [k]^{n-1}_\ell.
\label{eqn:HOMFLYPTbound}
\end{eqnarray}

Specializing to $k=2$, we can obtain a more explicit formula for the Jones polynomial.
The appropriate Markov weights $s^{(2,\ell)}_{[\lambda_1,\lambda_2]}$ when $\lambda_1 + \lambda_2 = n$ can be computed as follows.  By Lemma~3.5(b) of \cite{wenzl}, if $\lambda'$ is the $(2,\ell)$-diagram obtained by adding $r$ columns of two boxes to the left of another $(2,\ell)$-diagram $\lambda$, it follows that $s^{(2,\ell)}_{\lambda'} = [2]^{-2r}_\ell s^{(2,\ell)}_{\lambda}$.    A direct computation from the hook length formula (\ref{skl}) gives the Markov weight for a single row  diagram $[m]$ as
\[s^{(2,\ell)}_{[m]} = \frac{1}{[2]_\ell^{m}}\frac{[2]_\ell[3]_\ell\cdots [m+1]_\ell}{[1]_\ell[2]_\ell\cdots [m]_\ell} = \frac{[m+1]_\ell}{[2]_\ell^m}.\]
Since $\lambda = [\lambda_1,\lambda_2]$ can be obtained by adding $\lambda_2$ columns of two boxes to the left of the row diagram $[\lambda_1 - \lambda_2],$ it follows that
\[s^{(2,\ell)}_\lambda = \frac{s^{(2,\ell)}_{[\lambda_1 - \lambda_2]}}{[2]_\ell^{2\lambda_2}}
= \frac{[\lambda_1 - \lambda_2 + 1]_\ell}{[2]_\ell^{\lambda_1 + \lambda_2}}
=\frac{[\lambda_1 - \lambda_2 + 1]_\ell}{[2]_\ell^n}.\]
Plugging into (\ref{eq:HOMFLYPT}) thus leads to the following expression for the evaluation at $q = e^{2\pi i/\ell}$ of the Jones polynomial of the trace closure of a braid $b\in B_n$:
\begin{eqnarray}
J_{\h{b}}(e^{2\pi i/\ell}) &=& \overline{\tr}_{2\ell}(b) \nn \\
&=& \frac{1}{[2]_\ell} q^{-\frac 32 e(b)}
\sum_{\lambda \in \Lambda_n^{(2,\ell)}}
[\lambda_1 - \lambda_2 + 1]_\ell\Tr\pi^{(2,\ell)}_\lambda (b)
\nn \\
\end{eqnarray}
Because the numerator of the coefficient in front is just a complex phase, the absolute value of the Jones polynomial is
\begin{equation}
\big|J_{\h{b}}(e^{2\pi i/\ell})\big| = \big|\overline{\tr}_{2\ell}(b)\big| =  \frac{1}{[2]_\ell}\left|\sum_{\lambda\in \Lambda_n^{(2,\ell)}} [\lambda_1 - \lambda_2 + 1]_\ell \Tr\pi_\lambda^{(2,\ell)}(b)\right|. \label{absjonespoly}
\end{equation}

\subsection*{Jones polynomial of generalized closure from the Temperley-Lieb algebra}
Here we obtain representation-theoretic formulae
for the Jones polynomial of the generalized closure of a braid.  For this, we need to briefly describe the connection between the Jones polynomial and the Temperley-Lieb algebra.

\subsubsection*{The Temperley-Lieb algebra}
As we have described in the previous subsection,
the representations corresponding to the case $k=2$ are all that is relevant for the Jones
polynomial.  
In these cases, it is known that the representation matrices satisfy 
$\pi^{(2,\ell)}(e_i e_{i+1} e_i  -  \tau e_i) = 0$, where
$\tau = [2]^{-2}_\ell$. 
This can be shown using the fact that each of the $q$-deformed three-row anti-symmetrizers
\[1 - q^{-1}(g_i + g_{i+1}) + q^{-2}(g_ig_{i+1} + g_{i+1}g_i) - q^{-3}g_ig_{i+1}g_i\]
vanishes on representations with fewer than three rows (the proof is similar to the well-known undeformed case at $q=1$, where there can exist no totally antisymmetric state of three qubits).
The quotient of the Hecke algebra by the relations $e_ie_{i+1}e_i = \tau e_i$ yields the Temperley-Lieb algebra $TL_n(\tau)$ 
which is generated by 1 and idempotents $\{e_1,\dotsc,e_{n-1}\}$ satisfying
\begin{eqnarray}
e_i^2         & = & e_i  \label{tl1} \\
e_i e_{j} e_i & = & \tau e_i \quad\quad\,  |i-j|=1  \label{tl2} \\
e_i e_j       & = & e_j e_i  \hspace{.255in} |i-j|>1\,.  \label{tl3}
\end{eqnarray}
Temperley-Lieb algebra is advantageous because it provides representations of the  \emph{diagram monoid} $K_n$ \cite{kauffman1}, defined as follows.  For each $1\leq i\leq n-1$, define the ``cup-cap" diagram
\[\omega_i = \!\! \braid{1}{\POS (-3,.7)*{}; (-3,-.7) *{} **\dir{-}; \POS (-1,.7)*{}; (-1,-.7) *{} **\dir{-}; \POS (0,.7), \vcap[-1]; (0,-.7),\vcap[1]
\POS (2,.7)*{}; (2,-.7) *{} **\dir{-};
\POS (4,.7)*{}; (4,-.7) *{} **\dir{-};
\POS(-2.05,0) *{\cdots};
\POS(3.05,0) *{\cdots};
},\]
where the cup-caps act on strands $i$ and $i+1$.
Multiplication of cup-caps is performed pictorially (up to isotopy), amounting to stacking the generators $\om_i$ in the same way as with the generators of $B_n$.  Then, 
$K_n$ is generated by the cup-cap generators $\{\om_1,\dotsc,\om_{n-1}\}$, together with an extra closed-loop diagram $\delta$, which satisfy the relations
\begin{eqnarray}
\om_i^2         & = & \delta\om_i  \label{om1} \\
\om_i \om_{j} \om_i & = &  \om_i \quad\quad\, \hspace{.13in} |i-j|=1  \label{om2} \\
\om_i \om_j       & = & \om_j \om_i  \quad\quad |i-j|>1\,.  \label{om3}
\end{eqnarray}
These relations are perhaps best understood in terms of pictures which we invite the reader to draw, or otherwise to consult Figure~8 of \cite{kauffman1}.  While these relations are similar to those used above to define the Temperley-Lieb algebra, a more transparent connection is obtained by reformulating (\ref{tl1})--(\ref{tl3}) in terms of the scaled projections $E_i \equiv [2]_\ell e_i$, in which case the corresponding relations
\begin{eqnarray}
E_i^2         & = & [2]_\ell E_i  \label{TL1} \\
E_i E_{j} E_i & = & E_i \quad\quad\,  \hspace{.128in} |i-j|=1  \label{TL2} \\
E_i E_j       & = & E_j E_i  \quad\quad |i-j|>1\, \label{TL3}
\end{eqnarray}
are formally identical to (\ref{om1})--(\ref{om3}).
The diagram monoid $K_n$ is then represented inside $TL_n(\tau)$ via the map taking $\om_i \mapsto E_i$ and $\delta \mapsto [2]_\ell$.
As with $B_n$, we use this map to obtain representations of $K_n$ from those of $TL_n(\tau)$ by setting \[\pi^{(2,\ell)}_\lambda(\om_i) = \pi^{(2,\ell)}_\lambda(E_i) =
[2]_\ell\pi^{(2,\ell)}_\lambda(e_i).\]
For every even number $2p$,
define the rectangular tableau $t_{2p} \in T_{[p,p]}$ by
\newcommand{\nnn}{\mbox{ \!\!\raisebox{.02in}{\tiny $2p\!\!-\!\!1$}}}
\newcommand{\nnnn}{\mbox{$2p$}}
\begin{equation}t_{2n} = \rule{0in}{.2in}\,
{\Yvcentermath1\Yboxdim18pt\young(135,246)}\cdots
{\Yvcentermath1\Yboxdim18pt\young(\nnn,\nnnn)}\,\rule{0in}{.2in}.
\label{platvector}
\end{equation}
We slightly abuse notation and write the projection onto the corresponding vector $\ket{t_{2n}}$ as $t_{2p}\equiv \proj{t_{2p}}$.
We pause to mention a convention; given any $t\in T^{(2,\ell)}_{[p,p]}$ and $t'\in T_\lambda^{(2,\ell)}$, let $tt'$ be the tableau of shape $[p+\lambda_1,p+\lambda_2]$ obtained by adding $2p$ to each box of $t'$ and placing it to the right of $t$.  
It is easily checked that $tt' \in T_{[p+\lambda_1,p+\lambda_2]}^{(2,\ell)}$.  This defines a natural embedding of vector spaces 
$V_{[p,p]}\otimes V_{\lambda} \hookrightarrow V_{[p +\lambda_1,p+ \lambda_2]}$ which we denote $\ket{t}\ket{t'} \mapsto \ket{tt}$.  Throughout, we freely make this identification on vectors and operators.  
Now consider the following lemma.
\begin{lem} \label{lem:varphi2p}
Let $p$ and $n$ be positive integers satisfying $2p \leq n$.
If $\lambda = [\lambda_1, \lambda_2] \in \Lambda_n^{(2,\ell)}$, then
\[\pi_\lambda^{(2,\ell)}(\om_1\om_3\cdots \om_{2p-1}) =
\begin{cases}
[2]_\ell^p\,{t_{2p}}\otimes 1_{[\lambda_1 - p,\lambda_2 - p]} & \text{ if } \lambda_2 \geq p \\
0 & \text{ otherwise.}
\end{cases}\]
\end{lem}
\begin{proof}
Abbreviating $e_i = \pi_n^{(2,\ell)}(e_i)$, recall that $e_j\ket{t} = \ket{t}$ iff $j$ and $j+1$ are in the same column, while $e_j\ket{t} = 0$ iff $j$ and $j+1$ are in the same row.   Because the $e_{2i-1}$ are mutually commuting, if  $2i-1$ and $2i$ are in the same row of a given tableau for some $1\leq i\leq p$, $\ket{t}$ is annihilated by $e_1e_3\cdots e_{2p-1}$, and thus also by $\om_1\om_3\cdots \om_{2p-1}$.  Therefore, only vectors $\ket{t}$ for which $t = t_{2n} t'$ for some $t'\in T^{(2,\ell)}_{[\lambda_1-p,\lambda_2-p]}$ survive.  Clearly this is impossible if $\lambda_2 < p$.  On the other hand, for $1\leq i \leq p$ the $e_{2i-1}$, and therefore the $\om_{2i-1}$, may only act nontrivially on the leftmost $p$ columns of any given tableau.  By the convention mentioned above, this means that when $\lambda_2 \geq p$, the quantity in the lemma is proportional to the tensor product of the rank 1 projector $t_{2p}$ with an identity matrix as required.
\end{proof}

It is possible to combine $B_n$ and $K_n$ into a unified structure which was called a \emph{braid monoid} by Kauffman \cite{kauffman1}.  As this braid monoid contains diagrams consisting of both cup-caps and twists, its generators are the union of those of $B_n$ and $K_n$.  However, there are considerably more relations that these generators must satisfy (see e.g.\ \cite{bmw}) in addition to those of the braid group and diagram monoid.  For our purposes, however, we will not need the full braid monoid.  Instead, we shall only need to consider such generalized braids of the form $bw$, where $b\in B_n$ and $w\in K_n$ is a word in the $\om_i$'s, in which case $\pi^{(2,\ell)}_n(bw) = \pi^{(2,\ell)}_n(b)\pi^{(2,\ell)}_n(w)$.

\subsubsection*{Jones polynomial of the generalized closures}
For any braid $b\in B_n$, we give a formula for the absolute value of the Jones polynomial of the generalized closure of $b$ when $x=y=1_n\in B_n$.  Such a closure then depends only the parameters $p$ and $r$ which satisfy $2p + r = n$.  We obtain this formula by computing the normalized Markov trace of the braid $b$ with $p$ cup-caps at the tops of the leftmost $2p$ strands.  The formula is
\begin{eqnarray}
\big|J\big(\chi_{1,1}^{p,r}(b),e^{2\pi i/\ell}\big)\big| &=&
\left|\overline{\tr}_{2\ell}(b\om_1\om_3\cdots\om_{2p-1})\right| \nn\\
&=& \frac{1}{[2]_\ell}
\left|\sum_{\lambda \in \Lambda_n^{(2,\ell)}}
[\lambda_1 - \lambda_2 + 1]_\ell
\Tr\Big(\pi^{(2,\ell)}_\lambda(b)\pi^{(2,\ell)}_\lambda(\om_1\om_3\cdots\om_{2p-1})\Big)\right| \nn\\
&=& [2]_\ell^{p-1}
\left|\sum_{\lambda \in \Lambda_n^{(2,\ell)}:\lambda_2 \geq p}
[\lambda_1 - \lambda_2 + 1]_\ell
\Tr\Big(\pi^{(2,\ell)}_\lambda(b)({t_{2p}}\otimes 1_{\lambda - [p,p]})\Big)\right| \nn\\
&=& [2]_\ell^{p-1}\left|\sum_{\mu\in \Lambda_{r}^{(2,\ell)}} [\mu_1 - \mu_2 + 1]_\ell \Tr\Big(\pi_{[p,p]+\mu}^{(2,\ell)}(b)({t_{2p}}\otimes 1_{\mu}) \Big)\right|, \label{generalizedclosure}
\end{eqnarray}
In the second line, we have used the formula (\ref{absjonespoly}) for the absolute value of the normalized Markov trace.  The third line is by Lemma~\ref{lem:varphi2p}, while the last line is straightforward.

We remark here that this expression yields an immediate upper bound on the absolute value of the Jones polynomial of a link obtained as a generalized closure of a braid.  By unitarity, it is clear that the matrix traces above are maximized when the formula is evaluated at the identity braid $1_n\in B_n$.  In this case, we obtain 
\begin{eqnarray}
\big|J\big(\chi_{1,1}^{p,r}(1_n),e^{2\pi i/\ell}\big)\big| &=&
[2]_\ell^{p-1}\sum_{\mu\in \Lambda_{r}^{(2,\ell)}} [\mu_1 - \mu_2 + 1]_\ell \,d_\mu^{(2,\ell)} \nn\\
&=& [2]_\ell^{p -1} \sum_{\mu\in \Lambda_{r}^{(2,\ell)}} [2]_\ell^{r}\,s_\mu^{(2,\ell)}\, d_\mu^{(2,\ell)} \nn\\
&=& [2]_\ell^{p + r -1}, \label{maxjones}
\end{eqnarray}
where we have used (\ref{weightsdims1}) for the last step.
By specializing (\ref{generalizedclosure}) to the case where $r=0$ and $p=n$, we obtain
the following formula for the Jones polynomial of the plat closure of $b\in B_{2n}$:
\begin{eqnarray}
\left|J(\tilde{b},e^{i2\pi/\ell})\right|
&=& [2]_\ell^{n-1}\left| \Tr \big(\pi_{[n,n]}^{(2,\ell)}(b)\,t_{2n}\big)\right| \nn \\
&=& [2]_\ell^{n-1}\left|
\bra{t_{2n}}\pi_{[n,n]}^{(2,\ell)}(b)\ket{t_{2n}}\right|.
\label{jonesplat}
\end{eqnarray}

\Section{Algorithms for approximating the Jones and HOMFLYPT polynomials} \label{section:algorithms}
In this section, we use the circuits of Section~\ref{section:localqubitmodel} to give quantum algorithms for approximately evaluating, in the sense described in Section~\ref{section:links}, the Jones and one-variable HOMFLYPT polynomials at primitive roots of unity.  In particular, the results of this section constitute a proof of Theorem~\ref{thm:runningtime}.  We begin by giving a quantum algorithm which obtains an additive approximation of the one-variable HOMFLYPT polynomials $H^{(k)}$ of the trace closure of a braid, generalizing the Jones polynomial algorithm of \cite{ajl}.  If $k$ is a fixed constant, this algorithm runs in time polynomial in the length of the braid.  Then, we sketch two complementary ways in which this algorithm can be generalized to approximate the absolute value of the Jones polynomial of any generalized closure of a braid.

\subsection*{Approximating the HOMFLYPT of the trace closure}
Recall the setup of Problem~\ref{problem:approximatehomflypt}.  We are given a length-$m$ braid $b\in B_n$, positive integers $k,\ell$ satisfying $2\leq k < \ell$, and a number $\delta > 0$.  We assume that $k$ is a fixed constant.  Because a length $m$ braid can act nontrivially on at most $2m$ strands, we lose no generality in assuming that $n\leq 2m$.     
We showed in (\ref{eqn:HOMFLYPTbound}) that the maximal value of $\big|H^{(k)}_{\h{b}}(e^{2\pi i/\ell})\big|$ over all $b\in B_n$ is $[k]_\ell^{n-1}$.  We therefore focus on estimating the following normalized version of the one-variable HOMFLYPT which,  by  (\ref{eqn:HOMFLYPTtrace}) and (\ref{weightedtrace}), has the following expression for $b\in B_n$:
\begin{eqnarray*}
\frac{1}{[k]_\ell^{n-1}}H^{(k)}_{\h{b}}(e^{2\pi i/\ell}) &=& q^{\frac{k-1}{2}e(b)}\tr_{k,\ell}(b) \\
&=& e^{\frac{2\pi i(k-1)e(b)}{2\ell}}\sum_{\lambda \in \Lambda_n^{(k,\ell)}} s_\lambda^{(k,\ell)} \Tr\pi^{(k,\ell)}_\lambda(b).
\end{eqnarray*}
The Markov weights $s_\lambda^{(k,\ell)}$ are defined in (\ref{skl}).  We sample 
from a complex-valued random variable with the above expectation by a procedure similar to that of \cite{ajl}.  First observe that by (\ref{weightsdims1}), the numbers 
$\Big\{P_\lambda = s_\lambda^{(k,\ell)}\,d_\lambda^{(k,\ell)} : \lambda \in \Lambda_n^{(k,\ell)}\Big\}$ are probabilities.  Because $k$ is constant, the number of weights grows polynomially with $n$, so choosing a $\lambda\in \Lambda_n^{(k,\ell)}$ with probability $P_\lambda$ can be done efficiently because those probabilities can be precomputed and sampled from by a standard procedure.  Second, note that given $\lambda \in \Lambda_n^{(k,\ell)}$, there is an efficient procedure for choosing a path $t\in T_\lambda^{(k,\ell)}$ uniformly at random.  This is because the Young graph $\Lambda_{0,n}^{(k,\ell)}$ is a layered graph, with a polynomial number of diagrams $\leq \big(\min\{n,\ell - k\} + 1\big)^k$ in each layer.  Therefore, the number of paths from $\emptyset$ to a given node in the diagram can be computed in advance.  To choose a path ending in a particular diagram $\lambda\in \Lambda_n^{(k,\ell)}$, the trick is to move in reverse; starting at $\lambda$, choose the next diagram $\lambda'\in \Lambda_{n-1}^{(k,\ell)}$  with a probability proportional to the number of paths from $\emptyset$ to $\lambda'$, divided by the total number of paths from $\emptyset$ ending in the $n-1$'st layer. 

The algorithm proceeds as follows.  One first chooses a tableau $t\in T_n^{(k,\ell)}$ randomly as above.  Next, one samples from random variables $X,Y\in \{\pm 1\}$ with conditional expectations satisfying
\[\E[X + i Y | \lambda,t] = \bra{t}\pi_{\lambda}^{(k,\ell)}(b)\ket{t}.\]
This can be done efficiently using the quantum circuits for the Jones-Wenzl representation described in Section~\ref{section:localqubitmodel}, together with the following standard lemma which we prove in the appendix for convenience:
\begin{lem}[Sampling Lemma] \label{lemma:sampling}
Let $U$ be a quantum circuit of length $O\big(\poly(n)\big)$ acting on $n$ qubits, and let $\ket{\psi}$ be a pure state of the $n$ qubits which can be prepared in time $O\big(\poly(n)\big)$.
It is then possible to sample from random variables $X,Y\in \{\pm 1\}$
for which
\[\E [X + i Y] = \bra{\psi}U\ket{\psi}\]
in $O\big(\poly(n))$ time.
\end{lem}
This process is repeated, averaging the values of $X$ and $Y$, until the desired precision is met.  The nonconditional expected value of $X + i Y$ is 
\begin{eqnarray*}
\E[X + iY] &=& \sum_{\lambda\in \Lambda_n^{(k,\ell)}}  \frac{P_\lambda}{d_\lambda^{(k,\ell)}}
\sum_{t\in T_\lambda^{(k,\ell)}}\bE[X + iY|\lambda,t] \\
&=& \sum_{\lambda\in \Lambda_n^{(k,\ell)}} s^{(k,\ell)}_\lambda
\sum_{t\in T_\lambda^{(k,\ell)}} \bra{t}\pi_{\lambda}^{(k,\ell)}(b)\ket{t} \\
&=& \sum_{\lambda\in \Lambda_n^{(2,\ell)}} s^{(k,\ell)}_\lambda
\Tr\pi_{\lambda}^{(k,\ell)}(b).
\end{eqnarray*}
Therefore 
\[e^{\frac{2\pi i(k-1)e(b)}{2\ell}} \bE\big[X + iY] = \frac{1}{[k]_\ell^{n-1}}H^{(k)}_{\h{b}}(e^{2\pi i/\ell}),\]
so a sufficiently good estimate of $\bE\big[X + iY]$ provides the needed estimate of the normalized one-variable HOMFLYPT polynomial.  By the following lemma, we will see that the above procedure only needs to be repeated polynomially many times to obtain the desired estimate.  
\begin{lem}[Chernoff Bound]
Let $\{X_1,\dotsc,X_M\}$ be real-valued random variables satisfying $|X_j|\leq 1$ and
$\E X_j = \mu$. Then
\[\Pr\left\{\left|\frac{1}{M}\sum_{j=1}^M X_j - \mu\right| > \delta \right\} \leq 2\exp\left(-\frac{M\delta^2}{4}\right).\]
\end{lem}
Suppose the above algorithm is run $M$ times, obtaining random variables $\{X_j,Y_j\}_{j=1}^M$.  Setting $Z_j = X_j + iY_j$, the union bound and the above lemma imply 
\begin{eqnarray*}
\Pr\!\left\{\left|\frac{1}{M}\sum_{j=1}^M Z_j \!-\! \bE Z_1 \right| \!>\! \delta \right\} \!\!\!\!&\leq&\!\!\!\! 
\Pr\!\left\{\left|\frac{1}{M}\sum_{j=1}^M X_j \!-\! \bE X_1\right| \!>\! \frac{\delta}{\sqrt 2} \right\}
+ \Pr\!\left\{\left|\frac{1}{M}\sum_{j=1}^M Y_j \!-\! \bE Y_1\right| \!>\! \frac{\delta}{\sqrt 2} \right\} \\
&\leq& \!\!\!\!4\exp\left(-\frac{M\delta^2}{8}\right).
\end{eqnarray*}
Thus, for any $\delta>0$, the probability that $Z \equiv e^{\frac{2\pi i(k-1)e(b)}{2\ell}}\frac{1}{M}\sum Z_j$ deviates from the desired value of the normalized one-variable HOMFLYPT polynomial by more than $\delta$ is smaller than the desired error of $1/4$, provided that the number of samples satisfies $M > 32 \ln(2)/\delta^2$.  Since each sample is obtained in  $O\big(\poly(m)\big)$ time, the desired additive approximation is obtained in $O\big(\poly(m,1/\delta)\big)$ time as required.

\subsection*{Approximating the Jones polynomial of generalized closures}
In this section we show how to approximate the absolute value of the Jones polynomial of generalized closures of a braid.  As mentioned in Section~\ref{section:links}, we focus on the absolute value because there is no unambiguous way to assign an orientation to a generalized closure.  Recall the setup for Problem~\ref{problem:approximatejonesclosure}.  We are given a length-$m$ braid $b\in B_n$, 
two nonnegative integers $p,r$ for which $2p+r = n$ and a positive integer $\ell$.
We lose no generality in assuming that the top and bottom braids which define the generalized closure are $x,y=1_n$.  
As we showed in (\ref{maxjones}), the largest absolute value of the Jones polynomial is obtained when the closure yields $p+r$ trivial knots.  Therefore we focus on estimating the normalized quantity
\[\frac{1}{[2]_\ell^{p+r-1}}\left|J\big(\xi_{1,1}^{p,r}(b),e^{2\pi i/\ell}\big)\right|.\]
We show two different ways of achieving this goal, as each constitutes a different generalization of the previous algorithm which is interesting in its own right.  The first makes use of the following proposition \cite{jones87}, which shows that the generalized closure $\xi^{p,r}(b)$ can be written as the plat closure of a related braid on $2p+2r$ strands. 
\begin{prop}
Let $b\in B_{2p+r}$.  Then $\xi^{p,r}_{1,1}(b) = \widetilde{c^{-1} b c}$, where 
$c = \sig^{2p+2r-1}_{2p+r}\cdots \sig^{2p+5}_{2p+3}\sig^{2p+3}_{2p+2}$
and for $i < j$, we write $\sig_{i}^j = \sig_i\sig_{i+1}\cdots \sig_{j-2}\sig_{j-1}$ for the braid where strand $j$ is moved above its neighboring strands on the left and is inserted in position $i$.
\end{prop}

\begin{figure}
\[
\braid{2}{\caps{0}{1}\caps{2}{1}\caps{4}{1};
(0,2) \s\r\r\s; (0,1) \s\s\r\s\s;
(-.3,0) *{}="tl"; 
(2.3,0) *{}="tr"; 
(-.3,-2) *{}="bl";
(2.3,-2) *{} ="br"; 
"tl";"tr" **\dir{-};
"tl";"bl" **\dir{-};
"bl";"br" **\dir{-};
"tr";"br" **\dir{-};
(3,0); ; \s\s\s;
(3,-1); ; \s\s\s;
(0,-2); ; \s\s\l\s\s;
(0,-3); ; \s\l\l\s;
(0,-4); ; \cups{0}{1}\cups{2}{1}\cups{4}{1};
(1,-1) *{b};
(-.7,2) *{}="TL"; 
(5.4,2) *{}="TR"; 
(-.7,-4) *{}="BL";
(5.4,-4) *{} ="BR"; 
"TL";"TR" **\dir{--};
"TL";"BL" **\dir{--};
"BL";"BR" **\dir{--};
"TR";"BR" **\dir{--};
}
\]
\fcaption{\label{figure:tracefromplat}
The trace closure of $b\in B_3$ is isotopic to the plat closure of the modified braid
$ (\sig^5_3\sig^3_2)^{-1} b \sig^5_3\sig^3_2 = \sig_2^{-1}\sig^{-1}_4\sig^{-1}_3b\sig_3\sig_4\sig_2$.  Note that we identify $b$ with its image $\iota\circ\iota\circ\iota(b)\in B_6$.
}
\end{figure}
While we invite the reader to verify this, the main idea is contained in Figure~\ref{figure:tracefromplat}, where we illustrate how the trace closure can be obtained from the plat closure.  Note that since the braid $c$ in the proposition consists of $\sum_{i=1}^{t-1} i = O(t^2)$ generators, the modified braid $b'\equiv c^{-1}bc$ is at most polynomially longer than $b$ itself.
Setting $p' = p+r$, observe that since $b'\in B_{2p'}$, (\ref{jonesplat}) implies that
\begin{eqnarray*}
\frac{1}{[2]_\ell^{p+r-1}}\left|J\big(\xi_{1,1}^{p,r}(b),e^{2\pi i/\ell}\big)\right| &=&
\frac{1}{[2]_\ell^{p'-1}}\left|J\big(\widetilde{b'},e^{2\pi i/\ell}\big)\right| \\
&=& \left| \bra{t_{2p'}} \pi^{(2,\ell)}_{[p',p']}(b')\ket{t_{2p'}}\right|. 
\end{eqnarray*}
In the same manner as with the HOMFLYPT, 
one then samples from a complex random variable $Z$ satisfying 
$\bE Z = \bra{t_{2p'}} \pi^{(2,\ell)}_{[p',p']}(b')\ket{t_{2p'}}.$
After averaging the results a suitable number of times, the absolute value of the average is reported.  We omit the details as they are identical to the former case.

Another procedure for this task makes direct use of the following representation-theoretic formula for the absolute value of the Jones polynomial:
\begin{eqnarray}
\frac{1}{[2]_\ell^{p+r-1}}\left|J\big(\xi_{1,1}^{p,r}(b),e^{2\pi i/\ell}\big)\right|
&=& \frac{[2]_\ell^{p-1}}{[2]_\ell^{p+r-1}}
\left|\sum_{\lambda\in \Lambda_{r}^{(2,\ell)}} [\lambda_1 - \lambda_2 + 1]_\ell \Tr\Big(\pi_{[p,p]+\lambda}^{(2,\ell)}(b)({t_{2p}}\otimes 1_{\lambda}) \Big)\right| \nn\\
&=& \frac{1}{[2]_\ell^{r}}
\left|\sum_{\lambda\in \Lambda_{r}^{(2,\ell)}} [\lambda_1 - \lambda_2 + 1]_\ell \Tr\Big(\pi_{[p,p]+\lambda}^{(2,\ell)}(b)({t_{2p}}\otimes 1_{\lambda}) \Big)\right|. \label{normalizedgeneralized}
\end{eqnarray}
Recall that by (\ref{weightsdims1}), the numbers $\left\{P_\lambda = s_\lambda^{(2,\ell)} d_\lambda^{(2,\ell)}: \lambda \in \Lambda_r^{(2,\ell)}\right\}$ sum to unity and thus define probabilities. The algorithm proceeds as in \cite{ajl} by first 
selecting an irrep $\lambda\in \Lambda_{r}^{(2,\ell)}$ at random
with probability $P_\lambda \equiv s_\lambda^{(2,\ell)}d_\lambda^{(2,\ell)}$. Then, a path $t$ in the Young graph
$\Lambda_{\lambda}^{(2,\ell)}$ is chosen uniformly at random. 
By the same arguments as in the HOMFLYPT case, one may then sample from random variables $X,Y\in \{\pm 1\}$ with conditional expectations satisfying
\[\E[X + i Y | \lambda,t] = \bra{t_{2p}}\bra{t}\pi_{[p,p] + \lambda}^{(2,\ell)}(b)\ket{t_{2p}}\ket{t}.\]
This procedure is then repeated, averaging the results until the desired precision is attained, after which the absolute value is reported.  Using (\ref{normalizedgeneralized}), it is straightforward to show that 
\[\big|\!\E[X + i Y]\big| = \frac{1}{[2]_\ell^{p+r-1}}\left|J\big(\xi_{1,1}^{p,r}(b),e^{2\pi i/\ell}\big)\right|.\]
The remainder of the proof is identical to the HOMFLYPT situation.


\Section{Simulating quantum circuits with braids}\label{section:encoding}
In this section, we show how to ``compile" any quantum circuit on $n$
qubits consisting of $m$ two-qubit gates into a description of a braid
on $4n$ strands with $\poly(m)$ crossings in such a way that the
output of the circuit is encoded into an evaluation of the Jones
polynomial of the plat closure of the braid at any primitive $\ell$'th
root of unity with $\ell \geq 5, \ell\neq 6$.  While this was
first proved in \cite{flw1}, our proof is simpler as it only
uses the representation of $B_8$ corresponding to the rectangular
diagram $\Yboxdim{4pt}\yng(4,4)$.  

There are uncountably many quantum circuits on $n$ qubits, but countably many braids.  Therefore, we will have to settle for approximations of quantum circuits.  Given unitaries $U$  and $U'$ and some $0\leq \ep \leq 1$, we say that $U$  $\ep$-\emph{approximates} $U'$ if $\norm{U-U'}_\infty\leq \ep$, where for a square matrix $M$, $\norm{M}_\infty$ denotes the \emph{operator norm}, or largest singular value of $M$.  
Our main goal in this section will be to prove the following: 
\begin{theo}\label{thm:computingByBraids}
Let $U=U_{m} U_{m-1} \cdots U_1$ be a quantum circuit of length $m$
acting on $n$ qubits and let $\epsilon = \Omega\big(2^{-\poly(n)}\big)$.  Then there
is a braid $b\in B_{4n}$ with $O\big(\poly(m)\big)$
crossings such that $U$ is $\ep$-approximated by $\pi^{(2,\ell)}_{[2n,2n]}(b)$.
In this case, the following inequality holds:
\begin{equation}
\left|
|\bra{00\cdots 0} U \ket{00\cdots 0}|^2 -
\frac{1}{[2]_\ell^{2n-1}}\left|J(\tilde{b},e^{2\pi i/l})\right|^2
\right|
\leq
\epsilon\,.
\end{equation}
Moreover, such a braid can be found from a description of the circuit in
$O\big(\poly(m)\big)$ time on a classical computer.
\end{theo}

To prove Theorem~\ref{thm:computingByBraids} we must first introduce some machinery.  A finite set of gates in $U(\CH)$ is said to be
\emph{universal} if it generates a dense subset of $U(\CH)$.    The following lemma  implies that the measurement probabilities of an $\ep$-approximation of a quantum circuit differ by at most $\ep$ from those of the desired circuit.  The proof is in the appendix.
\begin{lem}  \label{lemma:distance}
Let $U$ and $U'$ be $d\times d$ unitaries for which $U'$ $\ep$-approximates $U$, where $\ep \leq 1$.  Then, for every pair of normalized pure states $\ket{\phi},\ket{\psi}\in \bC^{d}$, we have 
\begin{eqnarray*}
\left| |\bra{\phi} U \ket{\psi}|^2 - |\bra{\phi} U' \ket{\psi}|^2\right|
\leq \ep.
\end{eqnarray*}
\end{lem}
The next theorem assures that universal sets of unitaries suffice to efficiently $\ep$-approximate an arbitrary quantum circuit.
\begin{theo}[Solovay-Kitaev]
Let $\CG\subset U(d)$ be a finite universal set of unitaries on a Hilbert space of constant dimension $d$ which are closed under inverses, i.e.\ $\CG^{-1} = \CG$.  Then any unitary $U\in U(d)$ can be $\ep$-approximated by a sequence of $O\big(\polylog(1/\ep)\big)$ gates in $\CG$.  Moreover, such a sequence can be found by a classical computer in  $O\big(\polylog(1/\ep)\big)$ time.
\end{theo}

Fixing any finite, universal set of two-qubit unitaries $\CG_2\subset U(4)$, let $\CG_n\subset U(2^n)$ be the set of gates obtained by letting $\CG_2$ act on all pairs of neighboring qubits.
Given a length $m$ circuit $U_mU_{m-1}\cdots U_1$ consisting of gates from  $\CG_n$ and any \emph{fixed} desired degree of accuracy $\ep$, the Solovay-Kitaev Theorem further implies that for each $i$, there is an $\ep/m$ approximation $U'_i$ to $U_i$ which is obtained by a sequence of $O\big(\polylog(m)\big)$ gates in $\CG$, and also that such a sequence can be found in $O\big(\polylog(m)\big)$ time (since $\ep$ does not grow with $m$, it does not figure into the complexity estimate).  
Observe that for any unitaries $U$ and $U'$, the operator norm is \emph{stable}, meaning that $\norm{U - U'}_\infty = \norm{U\otimes 1_d - U'\otimes 1_d}_\infty$ for every finite dimension $d$.  Therefore, the same approximations hold for the embeddings of the local gates into $U(2^n)$.   By a telescoping sum and the triangle inequality,  we have
\begin{equation}
\norm{U_m\cdots U_2U_1-U'_m\cdots U'_2U'_1}_\infty \leq  \sum_{i=1}^m \norm{U_i - U'_i}_\infty.
\label{eqn:telescoping}
\end{equation}
Therefore, there is an $\ep$-approximation to the desired circuit which uses $O\big(m\,\polylog(m)\big)$ gates from $\CG_n$.
So we see that working with finite universal sets of gates is just as good as working with all possible two-qubit unitaries, provided that one can live with bounded errors and polylogarithmically more gates.  In Section~\ref{section:complexity}, we define classes of quantum computational problems which are insensitive to these limitations.

Throughout this section, we fix an $\ell \geq 5, \ell\neq 6$ and for each
diagram $\lambda$, we abbreviate $\pi_\lambda \equiv
\pi_\lambda^{(2,\ell)}$ and $V_\lambda = V_\lambda^{(2,\ell)}$.  We
begin by describing how to encode a single logical qubit into
$V_{\Yboxdim{3pt}\yng(2,2)}$.  This is a two-dimensional
representation of $B_4$ with orthonormal basis
\[\ket{t_0} \equiv \KET{\scriptsize \, \Yvcentermath1 \young(13,24)\,}  \hspace{.5in}
\ket{t_1} \equiv \KET{\scriptsize \, \Yvcentermath1 \young(12,34)\,}.\]
These basis states may also be viewed as paths in the Young graph (see Figure~\ref{qubitpaths}).
\begin{figure}
\begin{footnotesize}
\[
\xymatrix@R=.1cm{
& & {\yng(1,1)} \ar[rd]^1 \\
\emptyset \ar[r]^1 & {\yng(1)} \ar[ur]^2 \ar[dr]_1 & & {\yng(2,1)} \ar[r]^2 & {\yng(2,2)} \\
& & {\yng(2)} \ar[ru]_2
}
\]
\end{footnotesize}
\fcaption{\label{qubitpaths}Qubit path basis for $W_1$.  The upper path corresponds to
the basis state $\ket{t_0}$, the lower path to $\ket{t_1}$.}
\end{figure}
Let $t$ and $t'$ be $(2,\ell)$-tableaux of respective shapes $[j,j]$
and $[k,k]$.  Recall the convention mentioned before Lemma~\ref{lem:varphi2p} where we write $tt'$ for the tableau of shape $[j+k,j+k]$
obtained by adding $2j$ to each box in $t'$ then placing it to the
right of $t$.  Central to our proof is that $tt'$ will always be a
$(2,\ell)$-tableaux.  Recall that this allows us to define a natural embedding of vector spaces $V_{[j,j]}\otimes V_{[k,k]} \hookrightarrow V_{[j+k,j+k]}$ via the mapping $\ket{t}\ket{t'} \mapsto \ket{tt'}$.  
We encode $n$ qubits into
a subspace $W^{(n)}$ of the irreducible representation $V_{[2n,2n]}$
of $B_{4n}$ which is spanned by the $2^n$ computational basis states
$\big\{\ket{t_{x_1}t_{x_2}\cdots t_{x_n}} : (x_1,x_2,\dotsc,x_n) \in
\{0,1\}^n\big\}.$   By our embedding the subspace $W^{(n)}$ can be identified with a tensor
product $\bigotimes_{i=1}^nW_i$, where for each $1\leq i\leq n$, we
identify $W_i$ with $V_{\Yboxdim{3pt}\yng(2,2)}$, so that it is
spanned by $\ket{t_0}$ and $\ket{t_1}$.  In this subspace of
$V_{[2n,2n]}$, each consecutive group of four strands is associated to
a single qubit in the following way.  For each $1\leq j < k \leq n-1$,
define the subset $\Sig_{j}^k = \{\sig_j,\sig_{j+1},\dotsc,\sig_k\}$
of generators of $B_n$ and write $B_j^{k}\subseteq B_n$ for the
subgroup generated by $\Sig_j^{k-1}$.  For each $1\leq i \leq n-2$,
the image of the 8 strand braid group $B_{4(i-1)+1}^{4(i+1)}$ under
the representation $\pi_{[2n,2n]}$, when restricted to $W^{(n)}$, acts
nontrivially only on the subsystem $W_i\otimes W_{i+1}$.  In fact, it
rapidly approximates any $U \in SU(W_i\otimes W_{i+1})$.  To see this,
it suffices to look at the case $i=1$.  Note that $W_1\otimes W_2
\subset V_{\Yboxdim{3pt}\yng(4,4)}$.  The embedding
$V_{\Yboxdim{3pt}\yng(2,2)}\otimes
V_{\Yboxdim{3pt}\yng(2,2)}\hookrightarrow V_{\Yboxdim{3pt}\yng(4,4)}$
looks like
\begin{eqnarray*}
\ket{t_0}\ket{t_0} &\mapsto&
  \KET{\scriptsize \,\Yvcentermath1 \young(1357,2468)\,} \\
\ket{t_0}\ket{t_1} &\mapsto&
  \KET{\scriptsize \,\Yvcentermath1 \young(1356,2478)\,} \\
\ket{t_1}\ket{t_0} &\mapsto&
  \KET{\scriptsize \,\Yvcentermath1 \young(1257,3468)\,} \\
\ket{t_1}\ket{t_1} &\mapsto&
  \KET{\scriptsize \,\Yvcentermath1 \young(1256,3478)\,}.
\end{eqnarray*}
While this proves that $\dim V_{\Yboxdim{3pt}\yng(4,4)} \geq 4$, we
mention that its dimension is $13$ when $\ell=5$ and is $14$ when
$\ell > 6$.  So it is clear that $SU(W_1\otimes W_2)\subset
SU(V_{\Yboxdim{3pt}\yng(4,4)})$.  Consider now the following
proposition.
\begin{prop}[Freedman, Larsen, Wang \cite{flw2}]\label{prop:dense}
Let $\ell \geq 5, \ell \neq 6$.  The image of $B_{8}$ under $\pi^{(2,\ell)}_{\Yboxdim{3pt}\yng(4,4)}{}$
is dense in $SU\Big(V^{(2,\ell)}_{\Yboxdim{3pt}\yng(4,4)}\Big)$.
\end{prop}
With this density result in hand, we are now equipped to prove Theorem~\ref{thm:computingByBraids}. \\
\begin{proof}[Proof of Theorem~\ref{thm:computingByBraids}]
Set $\ep'=\ep/m = \Omega\big(2^{-m}/m)$.  Together with the
Solovay-Kitaev Theorem, Proposition~\ref{prop:dense}  implies 
that for each $U'\in SU(W_1\otimes W_2)$,
there is a braid $b\in B_8$
of length $O\big(\polylog(1/\ep')\big) = O\big(\poly(m)\big)$ for which
$\pi_{\Yboxdim{3pt}\yng(4,4)}(b)$ $\ep'$-approximates $U'$, where we
consider $U'$ to be embedded in $SU(V_{\Yboxdim{3pt}\yng(4,4)})$. Note
that while some unitaries 
$\pi_{\Yboxdim{3pt}\yng(4,4)}(\sig_i)$ will \emph{not} be contained in
$SU(W_1\otimes W_2)$, the
construction above ensures that $\pi_{\Yboxdim{3pt}\yng(4,4)}(b)$
can be made arbitrarily close to any $U'\in SU(W_1\otimes W_2)$.

Because of the four-periodicity of our encoding, the same arguments
apply for approximating gates on any other adjacent pair of qubits.  Now, let
$U=U_m U_{m-1}\cdots U_1$ as in the statement of the theorem. Without
loss of generality, we may assume that each $U_i$ has unit determinant.  By (\ref{eqn:telescoping}), we may conclude that 
there is a braid $b\in B_{4n}$ of length $O\big(\poly(m)\big)$ for which
$\pi_{[2n,2n]}(b)$ $\ep$-approximates the circuit $U$, and also that
such a braid can be determined on a classical computer in
$O\big(\poly(m)\big)$ time.  The rest of the theorem now follows by Lemma~\ref{lemma:distance} and the expression (\ref{jonesplat}) for the absolute value of the Jones polynomial of the plat closure; the normalization constant $1/[2]_\ell^{2n-1}$ is so because the braid has $4n$ strands.
\end{proof}

\Section{Complexity-theoretic applications of the Jones polynomial} \label{section:complexity}
Complexity theory characterizes the asymptotic consumption of resources, such as time and space, required to obtain the solutions of certain classes of problems on a computer.  Writing $\{0,1\}^*$ for the set of all binary strings of finite length, a \emph{decision problem} asks for the evaluation of some function $f\colon \{0,1\}^* \to \{0,1\}$ on an arbitrary input.  Such a function determines a \emph{language} $L\subseteq\{0,1\}^*$ defined as $L=f^{-1}(1)$, so equivalently, the task is to determine, for each bit string $x^n$, whether $x^n\in L$.
Examples of languages include the set of all binary strings of even parity and the set of prime numbers, written in binary.  A \emph{complexity class} is a collection of computational problems, usually defined in terms of the resources required to determine membership of arbitrarily long bit strings in each language contained in that class.  For instance, any class of languages is a complexity class.
Two well-known complexity classes are P and NP.  The former is defined to contain every language $L\subseteq\{0,1\}^*$ for which there is a uniform family $\big\{C_n\colon \{0,1\}^n\to \{0,1\}\big\}$ of polynomial-size classical circuits (see e.g.\ \cite{papadimitriou, preskillnotes} for a definition of a classical circuit) for which $C_n(x^n) = 1$ when $x^n\in L$, while $C_n(x^n) = 0$ otherwise.  By a \emph{uniform family}, it is meant that there should be a classical algorithm which, given the number $n$ as input, produces a description of the circuit $C_n$ in $O(\poly(n)\big)$-time.  We mention that this definition of a uniform family of circuits extends to quantum circuits in the obvious way.  
NP, on the other hand, contains those languages $L$ for which there is a uniform family of classical circuits $\big\{C_n\colon\{0,1\}^n\times \{0,1\}^{m}\to \{0,1\} \big\}$, where the number $m$ of \emph{guess bits} grows polynomially with $n$, so that for each $x^n\in L$, there exists a \emph{witness string}, or \emph{certificate} $y^{m}$, so that $C_n(x^n,y^{m}) = 1$ if $x^n\in L$, while if $x^n\notin L$, $C_n(x^n,y^{m}) = 0$ for 
every possible certificate $y^{m}$.  Below, we will recall the definitions of certain quantum generalizations of these classes, showing how they may be characterized using the Jones polynomial.  
 
The complexity class BQP is defined to contain those decision problems
whose solution can be determined with bounded error in polynomial time
in the \emph{standard quantum circuit model}.  Formally, a language
$L$ is in BQP if there exists a uniform family of quantum circuits $\{U_n\}$ of polynomial length such that 
\[|\bra{x^n00\cdots 0}U_n\ket{x^n00\cdots 0}|^2 
\begin{cases}
\geq 3/4 & \text{ if } x^n \in L \\
\leq 1/4 & \text{ if } x^n \notin L.
\end{cases}\]
This relates the usual definition of BQP, in which a single output qubit is measured to obtain the result of a computation, as follows.  Consider a circuit $U'_n$ such that if $x^n\in L$, a measurement of the first qubit of $U'_n\ket{x^n0\cdots0}$ in the computational basis yields 1 with probability $\geq 3/4$ while if $x^n \notin L$, that measurement outcome will be 0 with probability $\geq 3/4$.  Appending an additional qubit $\ket{0}^A$ and labeling the output qubit of $U'_n$ as $B$, we then take
\[U_n = (I^A\otimes U'_n) \big(\sig_x^A \otimes\proj{0}^B + I^A \otimes \proj{1}^B\big) (I^A\otimes U'^\dagger_n).\]

Note that the Solovay-Kitaev theorem implies that the class BQP
remains unchanged if the particular universal gate set chosen for
the model merely contains a universal set of gates for each pair of
qubits.   The following theorem essentially appears in \cite{bflw}; we prove it in this paper for the sake of completeness.

%
%

\setcounter{theo}{-1}
\begin{theo} \label{thm:PABQP}
Let $A$ be a random oracle which solves \textbf{\emph{Approximate Jones Closure}} with $\delta = 1/8$ for the plat closure and some constant $\ell \geq 5, \ell \neq 6$.  Then  $\text{\emph{P}}^A = $ \emph{BQP}.  Loosely speaking, this means that \textbf{\emph{Approximate Jones Closure}} is ``\emph{BQP}-complete."
\end{theo}
\begin{proof}
To begin, it is an immediate consequence of Theorem~\ref{thm:runningtime} that $\text{P}^A\subseteq \text{BQP}$.  To see that $\text{BQP} \subseteq \text{P}^A$, let $L\in \text{BQP}$ and let $U$ be $n$'th quantum circuit in the uniform family of circuits for $L$.   By definition, $U$ is of length $m=O\big(\poly(n)\big)$ and acts on $n' = O\big(\poly(n)\big)$ qubits.  For each $x^n$, define the circuit $W_{x^n} = \bigotimes_{i=1}^n X^{x_i}$, which acts as the identity on the ancillary qubits.
Because
\[\bra{00\cdots 0}W_{x^n}UW_{x^n}\ket{00\cdots 0} = \bra{x^n00\cdots 0}U\ket{x^n00\cdots 0},\]
it follows by Theorem~\ref{thm:computingByBraids} that for any constant $\ep > 0$, there is a braid $b$ of length $O\big(\poly(m)\big) = O\big(\poly(n)\big)$ on $4n'$ strands which satisfies
\begin{eqnarray*}
\left|\,\,\frac{1}{[2]_\ell^{2n'-1}}\left|J(\tilde{b},e^{2\pi i /\ell})\right|^2 -
|\bra{x^n00\cdots 0}U\ket{x^n00\cdots 0}|^2\right| 
&\leq& \ep.
\end{eqnarray*}
Moreover, there is an $O\big(\poly(n)\big)$-time classical algorithm for determining such a braid from a description of $W_{x^n}UW_{x^n}$.
By invoking the oracle $A$, we obtain a random variable $Z$ satisfying 
\[\Pr\left\{ \left|Z - \frac{1}{[2]_\ell^{2n'-1}}\Big|J(\tilde{b},e^{2\pi i/\ell})\Big|\,\right| < 1/8 \right\} > 3/4.\]
Since $|a^2 - b^2| \leq 2|a-b|$ for $0\leq a,b\leq 1$, the last two estimates can be combined via the triangle inequality to show that $\big|Z^2 - |\bra{x^n00\cdots 0}U\ket{x^n00\cdots 0}|^2\big| < 1/4 + \ep$ with probability at least 3/4. Choosing $\ep = 1/8$ is sufficient to complete the proof.
\end{proof}

%
%
The complexity class QCMA is a certain quantum analog of NP, consisting of those languages $L$
for which a computationally unbounded oracle (generally personified as
Merlin) can, with high probability, efficiently convince a
computationally bounded verifier (Arthur) that each $x^n\in L$ is
actually contained in $L$.  Cheating is also discouraged, in that if
$x^n\notin L$, the probability that Merlin will convince Arthur
otherwise should be small.  Formally, we say that a language $L$ is in 
QCMA if there is a uniform family of quantum circuits
$\{U_n\}$ of polynomial length such that for
each length $n$ string $x^n\in L$, there is a length $m$ binary string
$y^m$ for which $P(x^n,y^m)\equiv |\bra{x^ny^m00\cdots 0}U_n\ket{x^ny^m00\cdots 0}|^2 \geq 3/4$, while if $x^n \notin L$, we have $P(x^n,y^m) \leq 1/4$ for every such $y^m$.
We remark that the usual definition of QCMA (see e.g.\ \cite{an} for the original definition) is phrased in terms of measurement probabilities of a single output qubit; by the remarks following the definition above of BQP, the definition we give here is equivalent.
For the same reasons as with BQP, the exact value $3/4$ of the success probability is not important; what is important is that the square of the matrix element is \emph{strictly} bounded away from 1/2.

Given a complexity class C, a problem is said to be C-\emph{hard} if it is at least as hard as any other problem in the class C, in the sense that access to an oracle which immediately computes the solution of that problem allows the solution of any problem in the class C in polynomial-time.  If that problem is also contained in the class C, we say that it is C-\emph{complete}.  We will show that the following problem is QCMA-complete.

\begin{problem}[\textbf{Increase Jones Plat}]
Given a braid $b\in B_{2n}$ with $O\big(\poly(n)\big)$ crossings, a fixed integer $\ell\geq 5$, $\ell \neq 6$,
and a class of braids $\CC_{2n}\subset B_{2n}$ of length $O\big(\poly(n)\big)$ for which membership can be decided in $O\big(\poly(n)\big)$ time on a classical computer,
decide, with the promise that only these two cases can occur, whether there exists another braid $c\in \CC_{2n}$ for which the absolute value of the normalized Jones polynomial of the plat closure
\[\frac{1}{[2]_\ell^{n-1}}
\left|J(\widetilde{cbc^{-1}},e^{2\pi i/\ell})\right|^2 \geq 3/4
\]
or if it is $\leq 1/4$ for all braids $c\in \CC_{2n}$.
\end{problem}

\begin{theo} \label{thm:QCMA}
\textbf{\emph{Increase Jones Plat}} is \emph{QCMA}-complete.
\end{theo}
\begin{proof}
To begin, we demonstrate that \textbf{Increase Jones Plat} is in
QCMA. Given a braid $b\in B_{2n}$ of length $O\big(\poly(n)\big)$, Merlin begins by sending Arthur the classical description of the ``witness" braid $c$. Next, Arthur checks to see if  $c\in \CC_{2n}$, which can be done in polynomial-time by definition.
Because of the promise, Arthur only needs to learn the value of
\[
\frac{1}{[2]^{n-1}}
\left|J(\widetilde{c^{-1} b c},e^{2\pi i/\ell})\right|^2
\]
with an accuracy of $< 1/2$ with probability $\leq 3/4$.  For this, it is sufficient for him to run \textbf{Approximate Jones Closure} on a quantum computer with $\delta =1/4$, which takes $O\big(\poly(n)\big)$-time.

To see that this problem is QCMA-hard, and is thus QCMA-complete, suppose that $L\in$ QCMA. Let $U$ be the $n$'th circuit from the uniform family of quantum verifiers for $L$, and suppose that $U$ acts on $n'$ qubits while accepting witnesses of size $m$.  Letting $\ep>0$ be a constant to be determined at the end of the proof, Arthur precomputes a set of four braids $\CC_8\subset B_8$ which $\ep/m$-approximate the circuits
$\big\{X^{y_1}\otimes X^{y_2} : y_1,y_2\in \{0,1\}\big\}$ (obviously, there is nothing to compute when $y_1=y_2 = 0$).  This can be done in polynomial-time.  Define $\CC_{4n'}\subset B_{4n'}$ to consist of braids which act as the identity everywhere except for the strands $4n + 1, \dotsc 4n + 4m$, where only a single braid from $\CC_8$ can act on each group of eight strands $8i + 1, \dotsc, 8i + 8$ within those strands.   If we define $V_{y^m}$ in an analogous manner to $W_{x^n}$ in the proof of Theorem~\ref{thm:PABQP}, where it acts only by flipping the witness qubits, it follows by (\ref{eqn:telescoping}) that for each braid $c$ in $\CC_{4n'}$, the corresponding unitary $\pi_{[2n',2n']}^{(2,\ell)}(c)$ $\ep$-approximates $V_{y^m}$ for one and only one $y^m$ (this is because $\norm {V_{y^m} - V_{z^m}}_\infty = 2$ whenever $y^m\neq z^m$). 
Arthur then tells $x^n$ to Merlin, who sends him a witness braid $c$ from $\CC_{4n'}$.  Define $W_{x^n}$ as in the proof of Theorem~\ref{thm:PABQP}, except that it acts as the identity on the witness and ancillary qubits.  By the first half of Theorem~\ref{thm:computingByBraids}, Arthur may compute a braid $b$ for which 
the circuit $W_{x^n} U W_{x^n}$ is $\ep$-approximated by 
$\pi^{(2,\ell)}_{[2n',2n']}(b)$.  Therefore, we obtain by (\ref{eqn:telescoping}) that 
$\pi^{(2,\ell)}_{[2n',2n']}(c^{-1}bc)$ is a $3\ep$-approximation of $W_{x^n}V_{y^m} UW_{x^n}V_{y^m}$.  Since 
\[\bra{00\cdots 0}W_{x^n}V_{y^n} UW_{x^n}V_{y^n}\ket{00\cdots 0} =
\bra{x^ny^m00\cdots 0}U\ket{x^ny^m00\cdots 0},\] we may apply the rest of Theorem~\ref{thm:computingByBraids} to obtain that 
\[\left| \big|\bra{x^ny^m00\cdots 0}U \ket{x^ny^m00\cdots 0}\big|^2
- \frac{1}{[2]^{2n'-1}} \left|J(\widetilde{c^{-1} b c},e^{2\pi i/\ell})\right|^2 \,\,\right| \leq 3\ep.\]
In order to complete the proof, notice that because of the promise, it is sufficient to  choose $\ep = 1/8$.
\end{proof}
A rather large complexity class is PSPACE, which contains all languages which can be decided using only polynomial space.  In \cite{wocjan}, PSPACE is characterized as the class of languages which can be decided by applying the same polynomial-size circuit (possibly exponentially) many times.  Formally, it is shown there that
if $L\in$ PSPACE, then there exists a uniform family of polynomial-size quantum circuits $\{U_n\}$, together with a sequence of polynomial-time computable natural numbers $e_n$, for which, if $f\colon \{0,1\}^*\to \{0,1\}$ is such that $f^{-1}(1) = L$, then
\[\bra{x^n100\cdots 0}U_n^{e_n}\ket{x^n100\cdots 0} = f(x^n).\]
Using this characterization, we will show that the following problem is PSPACE-complete.
%
%
\begin{problem}[\textbf{Approximate Concatenated Jones Plat}]
Given a braid $b\in B_{2n}$ with $O\big(\poly(n)\big)$ crossings, a constant $\ell$ satisfying  $\ell \geq 5$, $\ell \neq 6$, and a positive integer $e$ which is describable by $O\big(\poly(n)\big)$ bits, decide, with the promise that only these two cases can occur, whether the squared absolute value of the normalized Jones polynomial of the plat closure of the concatenated braid $b^e$ satisfies
\[\frac{1}{[2]_\ell^{n-1}}\,
\left|J(\widetilde{b^e},e^{2\pi i/\ell})\right|^2 \geq 3/4\]
or if it is $\leq 1/4$.
\end{problem}
\begin{theo} \label{thm:PSPACE}
\textbf{\emph{Approximate Concatenated Jones Plat}} is \emph{PSPACE}-complete.
\end{theo}
\begin{proof}
To see that this problem is PSPACE-hard, choose any language
$L\in$ PSPACE, let $U$ be the $n$'th quantum circuit
from its associated uniform family and let $e=O\big(2^{\poly(n)}\big)$ be the corresponding exponent.  Suppose that $U$ acts on $n'$ qubits.
For a given $x^n$, we will show how to use an oracle for \textbf{Approximate Concatenated Jones Plat} to determine whether or not $x^n\in L$.  Define the circuit $W_{x^n}$ as in the proof of Theorem~\ref{thm:PABQP}, except that it also flips the $n+1$st qubit.
By choosing $\epsilon = \ep'/e = \Omega(2^{-\poly(n)})$ for a constant $\ep'$ to be determined later, we may obtain, via Theorem~\ref{thm:computingByBraids},
a braid $b$ with only $O\big(\poly(n))$ crossings such that
$\pi^{(2,\ell)}_{[2n',2n']}(b)$ $\ep$-approximates $W_{x^n}UW_{x^n}$.
Note that
\[f(x^n) = \bra{x^n100\cdots 0}(W_{x^n}UW_{x^n})^e\ket{x^n100\cdots 0} =
\bra{00\cdots 0}W_{x^n} U^eW_{x^n}\ket{00\cdots 0}.\]
Because $\ep'/e = \ep$, it follows by (\ref{eqn:telescoping}) that $\pi^{(2,\ell)}_{[2n',2n']}(b^e)$ $\ep'$-approximates the circuit $W_{x^n}U^eW_{x^n}$.  Choosing any $\ep' < 1/2$ is sufficient to complete the proof of PSPACE-hardness.
All that is left to do is to check that this problem is contained in PSPACE as well.  First, it is clear that \textbf{Approximate Concatenated Jones Plat} is contained in BQPSPACE, as our local qubit implementation of Section~\ref{section:localqubitmodel} requires only polynomial space.  On the other hand, Watrous has shown \cite{watrous2} that PSPACE $=$ BQPSPACE, completing the proof.   
\end{proof}
Another class of computational tasks is comprised of \emph{counting problems}, which ask, for a given function $f\colon \{0,1\}^*\to \{0,1\}$, ``how many $x^n\in \{0,1\}^n$ satisfy $f(x^n) = 1$?"  The class \#P contains those counting problems arising from a function $f$ for which $f^{-1}(1) \in$ P.  We refer the reader seeking more detail to \cite{papadimitriou} for a more rigorous definition.

%
%
\begin{theo} \label{thm:SP}
Given a braid $b\in B_{2n}$ with $O\big(\poly(n)\big)$ crossings and a fixed integer $\ell\geq 5$, $\ell \neq 6$, computing the $n$ most significant bits of the absolute value of the normalized Jones polynomial of the plat closure of $b$
\[\frac{1}{[2]_\ell^{n-1}}\left|J(\tilde{b},e^{2\pi i/\ell})\right|\]
is \#\emph{P}-hard.
\end{theo}
\begin{proof}
Suppose $f\colon \{0,1\}^n\to \{0,1\}$ is computable in polynomial-time on a classical computer.  It is well-known (see e.g. \cite{preskillnotes}), that there is a quantum circuit $U$, acting on $n'=O\big(\poly(n)\big)$ qubits,  which computes $f$ reversibly and exactly, namely
\[U\ket{x^n}\ket{y}\ket{00\cdots 0} =
  U\ket{x^n}\ket{y\oplus f(x^n)\oplus 1}\ket{00\cdots 0}.\]
The state
\[\ket{\psi} = \frac{1}{\sqrt{2^n}}\sum_{x^n\in \{0,1\}^n}\ket{x^n}\ket{1}\ket{00\cdots 0}\]
can clearly be prepared efficiently.  Observe that
\begin{eqnarray*}
\bra{\psi}U\ket{\psi} &=&
\frac{1}{2^n}\sum_{x^n} \braket{1}{f(x^n)} = \frac{N_1}{2^n},
\end{eqnarray*}
where $N_1 = |f^{-1}(1)|$ is the solution to the \#P problem of counting the solutions to $f$. Let $W$ be the unitary circuit which acts with Hadamard gates on the first $n$ qubits, flips the next qubit, while leaving the rest alone, and set $\ep = 4^{-n-2}$.  Clearly $W\ket{00\cdots 0} = \ket{\psi}$.   Since $|a-b| \leq \sqrt{|a^2 - b^2|}$ for $0\leq a,b\leq 1$,  it follows by Theorem~\ref{thm:computingByBraids} that, given a description of the circuit $W^{-1}UW$, we may, in polynomial-time, compute a braid $b$ for which
\begin{eqnarray*}\left| |\bra{00\cdots 0}W^{-1}UW\ket{00\cdots 0}| - \frac{1}{[2]^{2n'-1}}\left|J(\widetilde{b},e^{2\pi i/\ell})\right|\,\,\right|
\leq \sqrt{\ep} = 2^{-n-1}.
\end{eqnarray*}
In particular, this means that the $n$ most significant bits of the two terms in the above absolute value agree. Therefore, an oracle which gives the $4n'$ most significant bits of that evaluation of the Jones polynomial can be used to obtain all $n$ bits of the number $N_1$.
\end{proof}

In fact, $n$ bits is more than enough bits for \#P-hardness.  Learning just $\Omega(n^\delta)$ of the most significant bits, for any constant $\delta > 0$, is \#P-hard as well.  This can be seen by padding with a sufficiently large, but polynomial, number of extra qubits.  On the other hand, we will see that learning just the highest order bit of the Jones polynomial is PP-hard.

%
%
The complexity class PP, introduced in \cite{gill}, contains those languages recognized by some uniform family of polynomial-size probabilistic classical circuits with probability greater than 1/2.  Equivalently, $L$ is in PP if there is a nondeterministic polynomial-time machine which accepts $x^n\in L$ on more than half of its paths.  
\begin{theo} \label{thm:PP}
Given a braid $b\in B_{2n}$ with $O\big(\poly(n)\big)$ crossings and a fixed positive integer $\ell\geq 5$, $\ell \neq 6$, computing the most significant bit of the absolute value of the normalized Jones polynomial of the plat closure of $b$
\[\frac{1}{[2]_\ell^{n-1}}\left|J(\tilde{b},e^{2\pi i/\ell})\right|\]
is \emph{PP}-hard.
\end{theo}
\begin{proof}
The proof follows by exactly the same manner as in the proof of \#P-hardness.  The only difference is that the most significant bit of the Jones polynomial is sufficient for deciding if the number is solutions is greater than $2^{n-1}$.
\end{proof}

\Section{Conclusions} \label{section:conclusions}
We showed how to implement the unitary Jones-Wenzl representations of the braid group at a primitive $\ell$'th root of unity in polynomial-time on a quantum computer.  We constructed the image of each generator using a number of local two-qubit unitaries which was polynomial in the number $n$ of strands, independent of how $\ell$ grows with $n$.  We then used this model to give algorithms for obtaining additive approximations of one-variable HOMFLYPT polynomials $H^{(k)}$ on a quantum computer.  These algorithms run in time which is polynomial in the number of crossings and the inverse of the desired accuracy, independent of how fast $\ell$ grows with these parameters, provided that $k$ is a constant.  While our circuits implementing the Jones-Wenzl representations $\pi^{(k,\ell)}$ are of length $O\big(\poly(n,k)\big)$, it appears that the classical part of our algorithm causes the running time to scale exponentially with $k$.

Other authors have approached the problem of implementing the Jones representations on a quantum computer.  In \cite{kauffman3}, a representation of the three-strand braid group was given in a qubit architecture.  Another approach toward implementing the Jones representations is contained in \cite{sub}; however, the final conclusions of that paper rested on an incorrect assumption that unitary matrix elements could be measured precisely in a single time step.  Finally, a recent article \cite{gar} drew on a connection between topological quantum field theories and spin networks in order to outline a quantum algorithm for approximating the colored Jones polynomial; those authors left as an open question whether their algorithm could be efficiently implemented on a local quantum computer.  

On a classical computer, if the number of strands is constant (i.e.\ not included in the complexity estimate), then the Jones polynomial of the trace closure of a braid with $m$ crossings can be computed exactly by simply multiplying representation matrices of the braid group, requiring time which is polynomial in $m$.  However, accounting for the number of strands, the matrices will be exponentially large in the number $n$ of strands. On a quantum computer, however, multiplication of exponentially large matrices is not a problem, provided that they are sufficiently ``sparse," or rather, that they arise via local unitary transformations of the state space of the quantum computer.  On the other hand, there is a sense in which such large unitary matrices are not explicitly observable; an exponential number of queries seem to be required in order to learn the trace of the matrix representing the unitary evolution of a quantum circuit.  Nonetheless, the additive approximation of the Jones polynomial we achieved with a quantum algorithm is unobtainable on a classical computer, unless of course BPP $\neq$ BQP.  

On the other side of the fence, we have simplified the proof from \cite{flw1,flw2} which shows how to simulate the standard quantum circuit model with braids.  By using four strands per qubit, rather than the three of the original proof, we simplified the representation theory dramatically.  The four-periodicity of our encoding means that shifting the eight-strand braid which represents a given local unitary on two qubits by a multiple of four strands realizes the same gate on another pair of qubits.  The original result of \cite{flw1} requires consideration of two different cases, depending on whether the first qubit is located at an even or an odd site.  
Our encoding applies to schemes for topological quantum computation based on the braiding of nonabelian anyons with label ${\scriptsize \frac{1}{2}}$ in $SU(2)_{\ell -2}$ Chern-Simons field theories with $\ell \geq 5$, $\ell \neq 6$.  
On the other hand, our local qubit implementation of the Jones-Wenzl representations allows efficient simulation of the braiding of particles with label ${\scriptsize \frac{1}{2}}$ in $SU(k)_{\ell - k}$ Chern-Simons theories.  
While slightly less general than the results of \cite{fkw}, as the only mapping class group our methods apply to is the braid group, we have tried to make our presentation accessible to readers who are unfamiliar with the intricacies of topological quantum field theories. 

\Section{Acknowledgments}
We would like to thank Scott Aaronson, Stephen Bigelow, Andrew Childs, Matthias Christandl, John Preskill, Robert Raussendorff, Joost Slingerland and Zhenghan Wang for helpful discussions.  We also thank Michael Freedman for bringing the paper \cite{bflw} to our attention.  Some of our diagrams are based on examples from Aaron Lauda's 
\Xy-pic tutorial.
JY is thankful for support from the National Science Foundation under
grant PHY-0456720 through the Institute for Quantum Information at the
California Institute of Technology.
PW would also like to thank J\"{o}rg B\"{u}hler for discussions that helped to initiate this project.  He is also grateful for support by the National Security Agency under
Army Research Office contract number W911NF-05-1-0294.


\appendix

\noindent
\begin{proof}[Proof of Lemma~\ref{lemma:sampling}]
Introduce an extra system labeled $\CC$ which holds a single control qubit and denote the Hilbert space of the qubits acted on by $U$ by $\CA$.  Because $U$ consists of $O\big(\poly(n)\big)$ gates, then so does the controlled unitary $V\colon \CA\CC\to \CA\CC$, defined by $V\ket{\psi}\ket{0}  = \ket{\psi}\ket{0}$ and $\ket{\psi}\ket{1} = (U\ket{\psi})\ket{1}$.
Initialize the qubit in the state $\frac{1}{\sqrt{2}}\big(\ket{0} + \ket{1}\big)^\CC$, and prepare the state $\ket{\psi}^\CA$.  Applying the controlled unitary places everything into the state
\[\ket{\Psi}^{\CA\CC} = \frac{1}{\sqrt{2}}\big(\ket{\psi}\ket{0} + (U\ket{\psi})\ket{1}\big).\]
Writing $\psi = \proj{\psi}$,
the reduced density matrix $\tau^\CC$ of the extra qubit is then equal to
\begin{eqnarray*}
\tau^\CC &=& \Tr_\CA \proj{\Psi}^{\CC\CA} \\
&=& \frac{1}{2}\Tr_\CA \pmat{\psi & \psi U^\dagger \\ U \psi & U\psi U^\dagger}\\
&=& \frac{1}{2}\pmat{1 & \bra{\psi}U^\dagger\ket{\psi} \\  \bra{\psi}U\ket{\psi} & 1}\\
&=& \frac{1}{2}\Big(1_2 + \sig_X\text{Re}\bra{\psi}U\ket{\psi}  + \sig_Y\text{Im}\bra{\psi}U\ket{\psi}\Big).
\end{eqnarray*}
Since the $\pm 1$ eigenstates of $\sig_X$ are $\ket{\pm} = \frac{1}{2}\big(\ket{0}\pm \ket{1}\big)$, measuring $\sig_X$ on the qubit yields a classical random variable $X\in\{+1,-1\}$ for which
$\E X = \text{\rm Re}\bra{\psi}U\ket{\psi}$ as required. Similarly, by running the algorithm a second time, we can sample from $Y$, which satisfies $\E Y = \text{\rm Im}\bra{\psi}U\ket{\psi}$.
\end{proof}

\noindent
\begin{proof}[Proof of Lemma~\ref{lemma:distance}] 
By the triangle inequality,
\begin{eqnarray*}
\big|\bra{\phi}U\ket{\psi}\big|^2 &=& \big|\bra{\phi}U'\ket{\psi} + \bra{\phi}(U-U')\ket{\psi}\big|^2 
\leq \big|\bra{\phi}U'\ket{\psi}\big|^2 + \big|\bra{\phi}(U-U')\ket{\psi}\big|^2.
\end{eqnarray*}
The operator norm satisfies 
$\norm{M}_\infty = \max\{|\!\Tr MX | : \norm{X}_1 =1\}$, where 
the \emph{trace norm} $\norm{X}_1$ of a square matrix $X$ is defined to be the sum of its singular values.  Thus, 
\begin{eqnarray*}
\big|\bra{\phi}(U-U')\ket{\psi}\big|^2 
&=&  \big|\!\Tr (U-U')\ket{\psi}\bra{\phi}\big|^2 
\leq \Norm{U-U'}_\infty^2 \leq \ep^2 \leq \ep,
\end{eqnarray*}
Therefore, $\big|\bra{\phi}U\ket{\psi}\big|^2 - \big|\bra{\phi}U'\ket{\psi}\big|^2 \leq \ep.$ 
A similar argument gives the same bound with $U$ and $U'$ swapped, proving the lemma. 
\end{proof}

\end{document}